\documentclass[10pt, conference, compsocconf]{IEEEtran}

 \usepackage{amssymb}
\usepackage{graphicx,amsmath, amsfonts}
\usepackage[english]{babel}
\usepackage[utf8]{inputenc}
\usepackage[T1]{fontenc}
 \usepackage{pifont}
\usepackage{rotating}
\usepackage{color}
 \usepackage{bbm}
   \usepackage{stmaryrd}
   \usepackage{eufrak}
  \usepackage{marvosym}
 \usepackage{cite}
 \usepackage{tikz}
\newlength\shadedboxwidth
\long\def\shaded#1{\begin{trivlist}\item[]%
\setlength\fboxsep{2ex}% 
\setlength\fboxrule{.4pt}% 
\def\bgcolor{.98}% 
\setlength\shadedboxwidth\linewidth
\addtolength\shadedboxwidth{-2\fboxsep}%
\addtolength\shadedboxwidth{-2\fboxrule}%
\fcolorbox[gray]{0}{\bgcolor}%
{\parbox{\shadedboxwidth}{#1}}\end{trivlist}}
 
\newtheorem{numer}{\hspace{-4mm}($\spadesuit$\hspace{-1mm}}

\newtheorem{tnumer}{\hspace{-4mm}($\clubsuit$\hspace{-1mm}}

\newcommand{\eop}{\hfill {$\Box$}}

\newtheorem{theorem}{\hspace{-3.5mm}Theorem}
 
 \newtheorem{lemmaa}[theorem]{\hspace{-3.5mm}Lemma}
 
 \newtheorem{definitionn}[theorem]{\hspace{-3.5mm}Definition}
\newtheorem{observ}[theorem]{\hspace{-3.5mm}Observation}
\newtheorem{remark}[theorem]{\hspace{-3.5mm}Remark}

\newcommand{\theoremmm}[1]{\vspace{1.5mm}\begin{theorem} #1 \end{theorem} \vspace{2mm}}
\newcommand{\lemmaaa}[1]{\vspace{1.6mm}{\em \begin{lemmaa} #1 \end{lemmaa} }\vspace{2mm}}
\newcommand{\definitionnn}[1]{\vspace{1.6mm} {\em \begin{definitionn} #1 \end{definitionn} } \vspace{2mm}}
\newcommand{\observationnn}[1]{\vspace{1.2mm}\begin{observ} #1 \end{observ} \vspace{1.4mm}}

\newcommand{\spr}[2]{\text{\footnotesize\ding{72}}^{#1}_{\hspace{-0.5mm}#2}}
\newcommand{\spg}[2]{\text{\footnotesize\ding{73}}^{#1}_{\hspace{-0.5mm}#2}}

\newcommand{\fff}{\mathbbm{f}}

\newcommand{\oaa}{\curlywedgeuparrow }
\newcommand{\obb}{\curlyveeuparrow }

\newcommand{\oAA}{$ \mbox{$ \hspace{-1.5mm} \oaa \hspace{-3mm}\cdot\cdot \hspace{0.5mm}$}$}
\newcommand{\oBB}{$ \mbox{$ \hspace{-1.5mm} \obb \hspace{-3mm}\cdot\cdot \hspace{0.5mm}$}$}

\newcommand{\oaA}{\oaa  \hspace{-1.666mm}  $\raisebox{-0.55ex}{$\cdot$}$  \hspace{1.5mm}  }
\newcommand{\obB}{\obb  \hspace{-1.666mm}  $\raisebox{0.55ex}{$\cdot$}$  \hspace{1.5mm}  }

\newcommand{\rrestriction}{ \hspace{-1mm}\restriction \hspace{-1mm}    }

\newcommand{\Dd}{$\mathbb{D}$}

\newcommand{\Ddd}{\mathbb{D}}

\newcommand{\Tt}{{\cal{T}}}

\newcommand{\sep}{\mathfrak{T}}

\author{Tomasz Gogacz, Jerzy Marcinkowski\thanks{Supported by Polish National Science Centre grant DEC-2013/09/B/ST6/01535},  \\
Institute of Computer Science, University Of Wrocław,\\
}

\title{Red Spider Meets a Rainworm: Conjunctive Query Finite Determinacy Is Undecidable.}

\begin{document}
\maketitle

\begin{abstract}
We solve a well known and long-standing  open problem in database theory,  proving that Conjunctive Query Finite Determinacy Problem is undecidable.  
The technique we use  builds on the top of our Red Spider method 
which we developed in our paper [GM15] to show undecidability of the 
same problem in the ``unrestricted case'' -- when database instances are allowed to be infinite. We also show 
 a specific instance $Q_0$, ${\cal Q}= \{Q_1, Q_2, \ldots Q_k\}$ such that the set $\cal Q$ of CQs does not determine CQ $Q_0$ 
 but finitely determines it. Finally, we claim that while $Q_0$ is finitely determined by $\cal Q$, there is no FO-rewriting of $Q_0$, with respect to $\cal Q$, 
 and we outline a proof of this claim\footnote{This research was supported by Polish National Science Centre grant 2013/09/N/ST6/01188 (Tomasz Gogacz) and 
 by Polish National Science Centre grant DEC-2013/09/B/ST6/01535 (Jerzy Marcinkowski).}.
 \end{abstract}
% 

%%%%%%%%%%%%%%%%%%%%%%%%%%%%%%%%%%%%%%%%%%%%%%%%% NOWE KONIEC %%%%%%%%%%%%%%%%%%%%%%%%%%%%%%%%%%%%%%%%%%%%%%%%%%%%%%%%%%%%%%%%%%%

\section{Introduction\\}

{\em ``Assume that a set of derived relations is available in a stored form. Given a query, can it be computed from the derived relations and,
if so, how?''} is the first sentence of  [LY85]. 
Saying the same in today's language:

\shaded{
  The instance of the Conjunctive Query Finite  Determinacy Problem (CQfDP)
  problem is a set of conjunctive queries ${\cal Q}=\{Q_1,\ldots Q_k\}$, and 
another such query $Q_0$.

 The question is whether ${\cal Q}$ determines $Q_0$, which means that for each two database instances (that is {\bf finite} relational structures)
 ${\mathbb D}_1$ and ${\mathbb D}_2$ such that  
$Q({\mathbb D}_1)= Q({\mathbb D}_2)$ for each $Q\in \cal Q$, it also holds that  $Q_0({\mathbb D}_1) = Q_0({\mathbb D}_2)$.}

Answering queries using views appears in so many various contexts that it is indeed hard to imagine a more natural scenario in database theory. 
See for example 
[H01], or a recent thesis [F15] for a survey\footnote{Actually, [F15] does such a good job as a survey that we will stick to an absolutely  minimal
introduction here, especially that we have enough of new technical material to easily overrun  the page limit anyway.}.
The contexts where such scenario appears include for example  query optimization and caching 
 [DPT99],
or -- to see more recent examples -- [FG12] where the view update problem is studied and 
 [FKN13] where the context are description logics. In all the examples we mentioned so far we ``prefer'' $Q_0$ to be determined by $\cal Q$. 
 Yet another context, where it is ``preferred'' that $Q_0$ is not determined, is privacy: we would like to release some views of the database, but in a way that does not allow certain query to be computed. 
 
Many variants of the problem were being considered, and the case we study, where both the views and the query are defined by conjunctive queries,
and where the views we can see are ``exact''
is not the only possible (what we call ``determinacy'' is ``losslessness under the exact view assumption'' in the language of 
[CGLV00]).
Let us just cite the most recent results: [NSV10] shows that the problem is decidable if each  query from $\cal Q$ has only one free variable; in 
[A11] decidability is shown for $\cal Q$ and $Q_0$ being ''path queries''. This is generalized in [P11] to the the scenario where $\cal Q$ are path queries but $Q_0$ is any conjunctive
query. 
The paper  [NSV07] is  the first to 
present a negative result. It was shown there, that the problem is undecidable if unions of conjunctive queries are allowed
rather than CQs. In [NSV10] it was also proved that determinacy is undecidable if the elements of $\cal Q$ are CQs and $Q_0$ is a first order sentence 
(or the other way round). Another  negative result is presented in [FGZ12]: determinacy is shown there to be undecidable  if  $\cal Q$  is a DATALOG program and 
$Q_0$ is CQ. 

\subsection{Our contribution. Finite vs. unrestricted case.}

As we said, the case we study, where both the views and the query are CQs, is not the only one to be studied, 
but one of special importance, since  CQs --  as [NSV07] puts it -- are 
``the simplest and most common language to define views and queries''. 
The main technical result of this paper is:\smallskip

\begin{theorem}\label{main00} CQfDP is undecidable.\smallskip
\end{theorem}

 As usually in database theory one can consider two variants of the problem: {\em finite}, where all the
structures in question (which in our case means ${\mathbb D}_1$ and ${\mathbb D}_2$) are assumed to be finite, and {\em unrestricted}, where there is no
such assumption. Most of the results of  [NSV07],  [NSV10], [A11]  and [P11] that we report above hold true regardless of the finiteness assumption.

A theorem analogous to 
Theorem \ref{main00} but concerning the unrestricted Conjunctive Query Determinacy Problem (CQDP) is the main result of our earlier paper [GM15]. 
Since the unrestricted case is  viewed by the database theory community as less natural, it is fair to say that
 the result from [GM15] was perceived by the community as a step forward, but not  as one closing the problem (see e.g. [F15]). 

Since problems tend to be computationally harder in the finite case than in unrestricted, it was natural to conjecture, after [GM15], that
Theorem \ref{main00} should hold true. But its proof is significantly more difficult than the respective proof in [GM15]. Let us try to explain why it is so.

As the reader is going to see in Section \ref{g-r}, determinacy (both finite and unrestricted) boils down to the question,
whether some query (call it $red(Q_0)$)  
is true in {\bf all} structures  $\mathbb M$  (for the unrestricted case), or in all {\bf finite} structures $\mathbb M$  (for the finite case)
satisfying  ${\mathbb M} \models {\cal T}_{\cal Q}$ and ${\mathbb M} \models green(Q_0)$, where 
$green(Q_0)$ is a structure, depending  on $Q_0$,  and ${\cal T}_{\cal Q}$ is a set of tuple generating dependencies, depending  on $\cal Q$.

But in order to decide the above question for {\bf all} $\mathbb M$ it is enough to  study just {\bf one} {\em universal structure}, namely 
$chase({\cal T}_{\cal Q},green(Q_0))$. Determinacy holds if and only if $red(Q_0)$ is true in this single structure. 

This means that, when we encode some undecidable problem to show undecidability of (the unrestricted) CQDP, we only need to 
prove that whenever we start from a positive instance of our problem we get $\cal Q$ and $Q_0$ such that  $chase({\cal T}_{\cal Q},green(Q_0))\models red(Q_0)$, 
and whenever we 
start from a negative instance  we get $\cal Q$ and $Q_0$ such that  $chase({\cal T}_{\cal Q},green(Q_0))\not\models red(Q_0)$

No such  universal structure
exists for the finite case. This generates problems of two sorts: 

\textbullet~ How can we be sure (when we start from a negative instance\footnote{Notice that the roles of positive and negative instances have swapped. This is
due to the fact that finite determinacy is co-r.e. and unrestricted determinacy is r.e.})  
that ${\mathbb M} \models red(Q_0)$ is true in all relevant finite structures ${\mathbb M}$? Do we have any technique that is specific 
for finite models?

\textbullet~ In order to show (when we start from a positive instance) that  ${\mathbb M} \not\models red(Q_0)$, for some relevant ${\mathbb M}$,
we need to built a finite model, for a set of TGDs, which omits some conjunctive query. We are dangerously close here
to the issues related to finite controllability (see 
[R06], % Rosati
[BGO10], % Barany Gotllob Otto
[GM13]) % my. 
which are known to be difficult. 

\subsection{Our contribution. First Order non-rewritability}\label{malutka}

\noindent
Let, as always,  ${\cal Q}=\{Q_1\ldots Q_n\}$ be a  set of CQs and let $Q_0$ be a CQ.  
Then, by (slightly restated) definition, $\cal Q$ (finitely) determines $Q_0$ if 
there is a function $h^{Q_0}_{{\cal Q}}$ which, for a (finite) structure $\mathbb D$ over $\Sigma$,
takes, as its argument, the structure ${\cal Q}({\mathbb D})$ and returns ``yes'' or ``no'', depending on whether ${\mathbb D}\models Q_0$ or not.
Notice that ${\cal Q}({\mathbb D})$ is no longer a structure over $\Sigma$. Its signature consists of one $k$-ary relation symbol 
 for each  query $Q_i\in \cal Q$ having $k$ free variables. %We will call this relation symbol just $Q_i$.

It is known from [NSV07] that if $\cal Q$  determines $Q_0$ in the unrestricted sense then function $h^{Q_0}_{{\cal Q}}$ can 
be defined by a first order formula $\Theta^{Q_0}_{\cal Q}$, called {\em FO-rewriting} of ${Q}_0$ with respect to $\cal Q$: 
$h^{Q_0}_{\cal Q}({\cal Q}({\mathbb D}))=yes$ if and only if ${\cal Q}({\mathbb D})\models \Theta^{Q_0}_{\cal Q}$. The proof is via Craig Lemma, and 
since this lemma does not hold for First Order Logic over finite structures, it was natural to conjecture 
that the result will not survive if we restrict our attention to finite database instances. 
And indeed, as we are going to show:\smallskip

\begin{theorem}\label{nieprzepisywalnosc}
There are $\cal Q$  and $Q_0$ such that  $\cal Q$ finitely determines $Q_0$ but $h^{Q_0}_{{\cal Q}}$ is not first order definable.\smallskip
\end{theorem}

A detailed proof of this theorem is too long for a conference paper, and will only be presented in the full version of this paper. We however
outline it here, and all the important proof ideas are already present in this outline.

\section{Preliminaries}

We need nothing beyond some standard finite model theory/database theory notions. They are only recalled here 
in order to fix notations.

\subsection{Basic notions}\label{basic}

When we say ``structure'' we mean a relational structure $\mathbb  D$ over some signature $\Sigma$, i.e. a set of elements (vertices), 
denoted as $Dom({\mathbb D})$ (or just as $\mathbb D$ if no confusion is possible) and a set of relational atoms, whose arguments are elements of $\mathbb D$ and whose predicate names are from $\Sigma$.  Atoms are (of course) only positive. 
For an atomic formula $A$ (or for any other formula) we use notation ${\mathbb D}\models A$ to say that $A$ is true in $\mathbb D$. 

Apart from predicate symbols  $\Sigma$ can also contain constants. If $c$ is a constant from $\Sigma$ and $\mathbb D$ is a structure over $\Sigma$ then $c\in Dom({\mathbb D})$.

${\mathbb D}_1$ is a substructure
of $\mathbb  D$ (and $\mathbb  D$ is a superstructure of  ${\mathbb D}_1$) if for each atom $A$ if  ${\mathbb  D_1}\models A$ then ${\mathbb  D}\models A$. This implies that  $Dom({\mathbb  D_1})\subseteq Dom({\mathbb  D})$.

For two structures ${\mathbb  D}_1$ and  $\mathbb  D$ over the same signature $\Sigma$ a function $h: Dom({\mathbb  D_1})\rightarrow Dom({\mathbb  D})$   is called a homomorphism 
if for each $P\in \Sigma$ of arity $l$ and  each tuple $\bar a\in Dom({\mathbb D}_1)^l$ if ${\mathbb D_1}\models P(\bar a)$ then ${\mathbb D}\models P(h(\bar a))$ 
(where $h(\bar a)$ is a tuple of 
images of elements of $\bar a$). 

%Notice that ${\mathbb D}_1$ a substructure
%of $\mathbb  D$ if and only if identity is a homomorphism from ${\mathbb D}_1$ to $\mathbb  D$.

A conjunctive query (over $\Sigma$), in short CQ, is a conjunction of atomic formulas (over $\Sigma$) whose arguments are either variables or the constants from $\Sigma$, preceded by 
existential quantifier binding some of the variables. It is  important  to distinguish between a CQ and its quantifier-free part. 

%We usually write 
%$\Psi$ or $\Phi$ for a conjunction of atoms without quantifiers and $Q$ (possibly with a subscript) for conjunctive queries, so that we have something like:\vspace{-3mm}

%$$Q(\bar x) = \exists \bar y  \;\;\Psi(\bar y,\bar x)$$\vspace{-5mm}

%where  $\Psi(\bar y,\bar x)$ is a formula being a conjunction of atomic formulas and  $\bar x$ is a tuple of variables which are free in $Q$.

For a conjunction of atoms $\Psi$ (or for a CQ $Q(\bar x) = \exists \bar y \;\Psi(\bar y,\bar x)$) the canonical structure of $\Psi$, denoted as $A[\Psi]$, is the structure 
whose elements are all the variables and constants appearing in $\Psi$ and whose atoms are atoms of $\Psi$. It is useful to notice that 
for a finite structure $\mathbb D$ and a set $V\subseteq Dom({\mathbb D})$ there is a unique 
conjunctive query $Q$ such that ${\mathbb D}=A[Q]$ and that $V$
is the set of free variables of $Q$. 

For a CQ $Q(\bar x) = \exists \bar y \;\Psi(\bar y,\bar x)$ with $\bar x=x_1, \ldots x_l$, for a  structure $\mathbb D$ and for 
a tuple $a_1,\ldots a_l$ of elements of $\mathbb D$ 
we write ${\mathbb D}\models Q(a_1,\ldots a_l)$ when there exists a homomorphism $h: A[\Psi] \rightarrow \mathbb D$ such that $h(x_i)=a_i$ for each $i$. 

Sometimes we also write  ${\mathbb D}\models Q$. Then we assume that all the free variables of $Q$ are implicitly existentially quantified, so that the meaning of the notation is 
that there exists some  homomorphism $h: A[\Psi] \rightarrow \mathbb D$.

The {\bf most fundamental  definition} of this paper now, needed to formulate the problem we solve:
 for a CQ $Q$ and for a structure $\mathbb D$ by $Q({\mathbb D})$ we denote the ``view defined by $Q$ over $\mathbb D$'', which is the relation
$\{\bar a: {\mathbb D}\models Q(\bar a)\}$. 

\subsection{TGDs and how they act on a structure}\label{tgds}

A Tuple Generating Dependency (or TGD) is a formula of the form:\vspace{-8mm}

$$\forall \bar x, \bar y\; [\Phi(\bar x, \bar y)\Rightarrow \exists \bar z \; \Psi(\bar z,\bar y)]$$\vspace{-5mm}

where $\Psi$ and $\Phi$ are conjunctions of atomic formulas. The standard convention, which we usually obey, is that 
the universal quantifiers in front of the TGD are omitted.

From the point of view of this paper it is important to see a TGD -- let it be $T$, equal to $\Phi(\bar x, \bar y)\Rightarrow \exists \bar z \; \Psi(\bar z,\bar y)$ --
 as a procedure whose input is a structure $\cal D$ and whose output is a new structure 
being a superstructure of  $\cal D$:\smallskip

{\tt

find a tuple $\bar b$ (with 
% $|\bar a|= |\bar x|$ and
 $|\bar b|= |\bar y|$) such that:

\ding{172} ${\cal D}\models \exists \bar x\; \Phi(\bar x, \bar b)$ via homomorphism $h$ but

\ding{173} ${\cal D}\not\models \exists z\; \Psi(\bar z, \bar b)$;\smallskip

create a new copy  of $A[\Psi]$;\smallskip

output  ${\cal D}(T,\bar b)$ being a union of $\cal D$ and the new copy of $A[\Psi]$, with each $y$ from $A[\Psi]$  identified with $h(y)$ in $\cal D$.\medskip 
}

The message,  which will  be {\bf good to remember}, is that the interface between the ''new'' part of the structure, added by a single application of a TGD to 
a structure, and the ''old'' structure, are the free variables of the query in the right  hand side of the TGD.

\subsection{Chase} For a set $\cal T$ of TGDs and for a structure $\mathbb D$ let $chase_0({\cal T},{\mathbb D})={\mathbb D}$. Then, 
$chase_{i+1}({\cal T},{\mathbb D})$ is defined by the procedure:\smallskip

\noindent
${\cal D}:=chase_{i}({\cal T},{\mathbb D})$\\
{\bf forall} pairs $T$, $\bar b$, where $T$ is a TGD in $\cal T$ and $\bar b$ is a  tuple  
of elements of $chase_{i}({\cal T},{\mathbb D})$ {\bf do:}\\
%\hspace*{10mm}
$\{\;${\bf if} \ding{172} and \ding{173} hold in ${\cal D}$ for $\bar b$ and $T$
%\hspace*{12mm}
{\bf then} ${\cal D}:={\cal D}(T,\bar b)\;\}$;\\
$chase_{i+1}({\cal T},{\mathbb D}):=\cal D$.\smallskip

Then $chase({\cal T},{\mathbb D})$ is defined as $\bigcup_{i\in \mathbb N} chase_i({\cal T},{\mathbb D})$.
Notice that our chase is ``lazy'' -- we only produce new atoms and new elements  when needed.

\section{Outline of the Technical Part} Most of this paper is devoted to the proof of Theorem \ref{main00}.

Proving undecidability means encoding. Encoding means programming. Programming means the device that is being programmed, and the
programming language.

In [GM15] we developed the device. An elementary ``hardware'' object there is a structure called {\em spider} and an elementary 
``instruction'' able to act on a structure built out of spiders is a CQ called {\em spider query}.

In the Preliminaries we said that TGDs can ``act on a structure''. But how can a (conjunctive) query possibly do? 
This is explained in Section \ref{g-r}. 

In Section \ref{powolanie} we try -- without diving into details -- to define an interface between the device from [GM15] and 
the current paper. In Section \ref{climbing} we define a high level programming language to manipulate spiders. We also define
what is means to compile a program in such a language and show that this compilation is correct. This is new -- the problem 
we encoded in [GM15] was much simpler than what we are going to encode here, and -- using our running metaphor -- we could 
afford programming in a language which was pretty low level. Some ideas of the proof of Lemma \ref{o-kompilacji} were however present 
already in [GM15].

{\bf Separating example}
As we already have noticed, finite determinacy is co-r.e. and unrestricted determinacy is r.e. which implies -- since we know from 
[GM15] that unrestricted determinacy is undecidable --  that the two notions do not coincide. But no separating example was know so far.
In Section \ref{sep-example} we construct such an example. This is not just for curiosity. As it appears, this example is (together with Rainworm Machines) one of the two main engines 
of the proof of Theorem \ref{main00} which is presented in Section \ref{najwazniejsza}.

Finally, in Section \ref{outline} we explain the main ideas of the proof of Theorem \ref{nieprzepisywalnosc}. The separating example from Section \ref{sep-example}
turns out to also be a counterexample for 
FO-rewritability.

\section{Green-Red TGDs}\label{g-r}

\subsection{\sc Green-Red Signature}

For a given signature $\Sigma$ let $\Sigma_G$ and $\Sigma_R$ be two  copies of $\Sigma$ with new relation symbols, which have the same names and the same arities as 
symbols in $\Sigma$ but are written in green and red respectively. Let $\bar\Sigma$ be the union of   $\Sigma_G$ and $\Sigma_R$.
Notice that the constants from $\Sigma$ (if there are any) are not relation symbols, so they are never colored and thus survive in $\bar\Sigma$ unharmed. 
 
For any formula $\Psi$ over $\Sigma$ let $R(\Psi)$ (or $G(\Psi)$) be the result of painting all the predicates in $\Psi$ red (green). 
For any formula $\Psi$ over $\bar\Sigma$ let $dalt(\Psi)$ (''daltonisation of $\Psi$'') be a formula over $\Sigma$ being  the result of erasing the colors from 
predicates of $\Psi$. The same convention applies to structures. 
%Whenever an uncolored relation symbol (usually {\small H}) is used 
%in the context of  $\bar \Sigma$  it should be understood as ''G({\small H}) or R({\small H})''. 
For a structure $\cal D$ over $\bar\Sigma$, by ${\cal D}\rrestriction G$ we mean the substructure of $\cal D$ consisting of all
its atoms which are over $\Sigma_G$. Analogously for 
${\cal D}\rrestriction R$.

\subsection{\sc Having ${\mathbb D}$ instead of ${\mathbb D}_1$ and  ${\mathbb D}_2$.}\label{trzy-be}

We  restate CQfDP a little bit, as we prefer  to be talking about one two-colored database instead of two. 
Clearly CQfDP can be equivalently restated as {\bf CQfDP.2}:\smallskip

\shaded{
\noindent
 The instance of the problem is a finite 
set $\cal Q$  of conjunctive queries 
and 
another conjunctive query $Q_0$, all of them over some signature $\Sigma$. 
The question is whether for each finite structure ${\mathbb D}$ over  $\bar\Sigma$ such that:\smallskip

\ding{182} $(G(Q))({\mathbb D})= (R(Q))({\mathbb D})$ for each $Q\in \cal Q$\smallskip

it also holds that $(G(Q_0))({\mathbb D})= (R(Q_0))({\mathbb D})$.}\medskip

\definitionnn{\label{g-strzalka-r}
For a conjunctive query $Q$ of the form $\exists \bar x \; \Phi(\bar x,\bar y)$ where $\Phi$ is a conjunction of atoms over $\Sigma$ let 
$Q^{G\rightarrow R}$  be the TGD {\em generated by} $Q$ in the following sense:\vspace{-4mm}

 $$Q^{G\rightarrow R}\;\;\;=\;\;\;\forall \bar x, \bar y \;[\; G(\Phi)(\bar x,\bar y) \Rightarrow \exists \bar z \; R(\Phi)(\bar z,\bar y)\;]$$\vspace{-4mm}
}

TGD $Q^{R\rightarrow G}$ is defined in an analogous way. For a set $\cal Q$ as above let ${\cal T}_{\cal Q}$ be the 
set of all TGDs of the form $Q^{G\rightarrow R}$ or $Q^{R\rightarrow G}$ with $Q\in \cal Q$.
It is very easy to see that:

\lemmaaa{
The above condition \ding{182} is satisfied by structure $\mathbb D$ if and only if ${\mathbb D}\models {\cal T}_{\cal Q}$.
}

Now  CQfDP.2 can be again restated as {\bf CQfDP.3}:

\shaded{
Given a set $\cal Q$ (as in the formulation of CQfDP.2 above), and 
another conjunctive query $Q_0$, 
 is it true that:\smallskip

\ding{183} for  each finite structure $\mathbb D$ and each tuple $\bar a$ of elements of $\mathbb D$,
if ${\mathbb D}\models {\cal T}_{\cal Q}, G(Q_0)(\bar a)$ then  also  ${\mathbb D}\models R(Q_0)(\bar a)$\hfill ?}

Of course the unrestricted version of CQfDP (called CQDP, see the Introduction) is equivalent to CQfDP.3 after removing, from its formulation,
the word ``finite''.

The equivalent version of Theorem \ref{main00} which we actually prove in Sections \ref{sep-example} and \ref {najwazniejsza} is:\medskip

\begin{theorem}\label{main1b}
%[Theorem \ref{main00} restated]
{\bf CQfDP.3} is undecidable.\medskip
\end{theorem}

%%%%%%%%%%%%%%%%%%%%%%%%%%%%%%%%%%%%%%%%%%%%%%%%%%%%%%%%%%%%%%%%%%%%%%%%%%%%%%%%%%%%%%%%%%%%%%%%%%%%%%%%%%%%%%%%%%%%%%%%%%%%%%%%%%%
%%%%%%%%%%%%%%%%%%%%%%%%%%%%%%%% %%%%%%%%%%%%%%%%%%%%%%%%%%

\section{Building on top of [GM15]: spiders and spider queries.}\label{powolanie}

\subsection{\sc \sc How to read this paper}

This paper builds on top of the techniques developed in [GM15].  
But we are of course  not able to include here a presentation of the techniques from [GM15] which would be 
detailed enough to make the current paper self-contained.
There are two (or three) possible ways of reading this paper:

\textbullet ~ A good way is to first read Sections IV, V and VI.A of [GM15] (which is about 4 pages) and then jump to Section \ref{climbing} of the current paper.

\textbullet ~ The shortest way is to read the next subsection, where we try to outline the ideas from [GM15] without going into details. 
We believe that most of the constructions of this paper should be understandable then, with the exception of the proof of Lemma \ref{o-kompilacji}(1), which constitutes
an interface between the techniques from [GM15] and the new material. 

\textbullet ~ It is of course not at all forbidden to read both the next subsection  and Sections IV, V and VI.A of [GM15] in any chosen order. 

\subsection{\sc {\sc Spiders and spider queries -- a crash course}\label{crash}}

As we observed in Section \ref{g-r}, conjunctive query (finite) determinacy is about
(finite) structures being models of $\cal T_Q$, for some set $\cal Q$ of CQs. So negative results concerning determinacy 
(in its both versions) are about writing nontrivial logic programs in the language having $\cal T_Q$ as its set of instructions. 
 $\cal T_Q$ are TGDs, and normally encoding complicated things using TGDs should not be that difficult. But the TGDs from  $\cal T_Q$
 are of a very special form, with right-hand side being merely an opposite-color version of the left-hand side. To understand to what extent this is a restriction 
 it is  good to realize that:
 
 \observationnn{[Very easy] Let $\mathbb D$ be a structure over $\Sigma_G$ and let $\cal Q$ be a set of CQs. Then there exists a 
 homomorphism $h: dalt(chase({\cal T_Q},{\mathbb D}))\rightarrow dalt({\mathbb D})$. 
 }
 
 This Observation seems to imply that  ${\cal T_Q}$ cannot produce anything ``non-trivial'' -- we never get,
 in $dalt(chase({\cal T_Q},{\mathbb D}))$, anything we did not have in $\mathbb D$ anyway\footnote
 {When we observed this phenomenon we were pretty sure we had just discovered a key fact in a decidability proof
 .}.
 
But, as we have discovered in [GM15], while daltonisation of $chase({\cal T_Q},{\mathbb D}$) is indeed always an uncomplicated structure, there is a lot we 
can gain by playing with colors. 

Let $s$ be a natural number, large enough, and let $\mathbb S=\{1,2,\ldots s\}$.
We defined a structure over $\Sigma$, which we call {\em spider}\footnote{
Spider is parametrized by $s\in \mathbb N$, so in [GM15] we call it $s$-pider.} 
 and colored versions of spider: $\spg{}{}$  and $\spr{}{}$ (green and red spiders). Each 
 spider has $2s$ legs, including $s$ ``upper'' and $s$ ``lower'' legs,
 and $\spg{I}{J}$ is a green spider with his $I$-th  upper leg and $J$-th lower leg being red (analogously for $\spr{I}{J}$). 
 $I,J\subseteq \mathbb S$  {\bf are always} either
 singletons or empty, so we can have green spiders with none, one or two red legs\footnote{$\spg{}{}$ is short for $\spg{\emptyset}{\emptyset}$. } (and {\em vice versa}). 
 The set of all spiders\footnote{We should distinguish here between $2+4s+2s^2$ ``ideal spiders'', which are elements of $\mathbb A$, and
 ``real spiders'' -- homomorphic copies  of the elements of $\mathbb A$ in some bigger structure.}
 of the form $\spg{I}{J}$ or  $\spr{I}{J}$ is denoted as $\mathbb A$.
 
 Then we have the set ${\mathbb F}$ of  {\bf spider queries}. Elements of ${\mathbb F}$ are denoted as $\fff^I_J$, where $I$ and $J$ are as always. The crucial
 observation is that the left hand side of the TGD  $(\fff^I_J)^{R\rightarrow G}$ matches with $\spr{I'}{J'}$ if and only if $I'\subseteq I$ and $J'\subseteq J$,
 and what is produced\footnote{``Matches'' means in particular, that the canonical structure of 
 the query $\fff^I_J$ is a substructure of the respective spider. 
 ``Produced'' means that $\spg{I\setminus I'}{J\setminus J'}$ emerges somehow in the structure, after the new vertices and new edges are added, 
 as demanded by the right-hand side of the TGD
 $(\fff^I_J)^{R\rightarrow G}$. Notice that all the new edges are green, so if there are any red edges in the resulting $\spg{I\setminus I'}{J\setminus J'}$
 they are inherited from the old structure.}
 by such an application of the TGD is $\spg{I\setminus I'}{J\setminus J'}$. The same of course holds for the colors reversed. 
 We like this trick so much that we call it the Rule of Spider Algebra:\\
 \hspace*{27mm} $\fff^I_J(\spr{I'}{J'})=\spg{I\setminus I'}{J\setminus J'}$ \hfill $(\clubsuit)$\medskip
 
  {\bf Binary queries}. 
 One more feature of a spider is that it has two  vertices not involved in the mechanism enforcing  the rule $\clubsuit$. 
 We call them {\em antenna} and {\em tail}. Also each query $\fff^I_J$ has its antenna and its tail. Their role is to form -- as we call them -- {\em binary queries}.

 The set ${\mathbb F}^2$ of   binary queries contains, for each two queries $\fff^I_J$ and $\fff^{I'}_{J'}$ two conjunctive queries
 $\fff^I_J\oaa \fff^{I'}_{J'}$ and $\fff^I_J\obb \fff^{I'}_{J'}$. 
 
 To define a conjunctive query one needs to specify its canonical structure and its set of free variables (which is a subset of the set of vertices of the canonical structure). The canonical structure of $\fff^I_J\oaa \fff^{I'}_{J'}$ is the disjoint union of canonical structures of $\fff^I_J$ and of $\fff^{I'}_{J'}$, with 
 the only exception that the antennas of $\fff^I_J$ and of $\fff^{I'}_{J'}$ are identified. Their joint antennas are/is an existentially 
 quantified  variable in $\fff^I_J\oaa \fff^{I'}_{J'}$,
 and their tails are free variables in $\fff^I_J\oaa \fff^{I'}_{J'}$. All the remaining variables are free if and only if they were free in 
 $\fff^I_J$ or  $\fff^{I'}_{J'}$ -- they do the magic of $\clubsuit$ that we do not want to go into. 
 
 Now please come back to Definition \ref{g-strzalka-r} and notice that, when a  query $Q= \fff^I_J\oaa \fff^{I'}_{J'}$ 
is seen as a green-red TGD $Q^{G\rightarrow R}$, the free variables in $Q$ are what connects the new part of the structure, added by a single execution of 
 $Q^{G\rightarrow R}$, to the old part. So, what a single execution of $Q^{G\rightarrow R}$ does is as follows: it finds, in the current structure,
 two green (real)  spiders (call them ${\cal S}$ and ${\cal S}'$) with tails $a$ and $a'$, which share their antennas, 
 and such that, according to $\clubsuit$,  $\fff^I_J$ can be applied to
 ${\cal S}$ and $\fff^{I'}_{J'}$ can be applied to ${\cal S}'$. Then it creates two new (real) red spiders $\fff^I_J({\cal S})$ and 
 $\fff^{I'}_{J'}({\cal S}')$, which again share the antennas, and their shared antenna is a new vertex. They are connected to the old structure 
 via $a$, which is also the tail of the new $\fff^I_J({\cal S})$, via $a'$ which is also the tail of the new  $\fff^{I'}_{J'}({\cal S}')$ and
 via some vertices of the old structure which correspond to other (than the two tails) free variables of $\fff^I_J\oaa \fff^{I'}_{J'}$.
 
What concerns $\fff^I_J\obb \fff^{I'}_{J'}$, again its canonical structure is the disjoint union of canonical structures of $\fff^I_J$ and of $\fff^{I'}_{J'}$.
The only difference is that now the tails of $\fff^I_J$ and of $\fff^{I'}_{J'}$ are identified as one variable, and this variable is 
 existentially 
 quantified  in $\fff^I_J\obb \fff^{I'}_{J'}$. 
 The two antennas are now free variables. The way $(\fff^I_J\obb \fff^{I'}_{J'})^{G\rightarrow R}$ acts on a structure is
 analogous to the one of $(\fff^I_J\oaa \fff^{I'}_{J'})^{G\rightarrow R}$, which was explained above. Same for $(\fff^I_J\oaa \fff^{I'}_{J'})^{R\rightarrow G}$
 and $(\fff^I_J\obb \fff^{I'}_{J'})^{R\rightarrow G}$.

%% CLIMBING ABSTRACTION LADDER %%
\section{Climbing the abstraction ladder}\label{climbing}

The language of spiders, whose signature is  $\bar\Sigma$, which we briefly described in the previous section,
is a very low level one -- we think it is Abstraction Level Zero. 
But in order to show our undecidability result we need to produce pretty complicated
programs in the language of spider queries from ${\mathbb F}^2$. 
The typical Computer Science way in such situation is to define a more abstract, higher order  language, use it as
the actual programming language (so that one does not need to worry about low level implementation details), and then compile 
the program written in this higher order language into the executable low-level form

Defining such a higher order language (or rather languages -- we will first {\em precompile} the original program into an intermediate language and then 
 {\em compile})
is precisely what we are going to do in this section. Each of the two languages we are going to define will comprise a relational signature and  a set of graph rewriting rules
(being TGDs,  in disguise) which will act on structures over this signature.\medskip

\noindent
{\bf Abstraction Level 1 language.} The signature  consists of one binary relation  ${\text\small H}({\cal S},\_,\_)$ for each (ideal) spider ${\cal S}\in \mathbb A$.
A structure over this signature will be called a {\em swarm}. Now we are going to define the set $\mathbb{L}_1$ of {\em swarm rewriting rules}.

\definitionnn{
For each query $\fff^{I_1}_{J_1}\obb \fff^{I_2}_{J_2}$ from ${\mathbb F}^2$  (and for each query $\fff^{I_1}_{J_1}\oaa \fff^{I_2}_{J_2}$  from ${\mathbb F}^2$)   
there will be a rule $\fff^{I_1}_{J_1}\obB \fff^{I_2}_{J_2}$ (resp.  $\fff^{I_1}_{J_1}\oaA \fff^{I_2}_{J_2}$) in ${\mathbb L}_1$, where 
 $\fff^{I_1}_{J_1}\obB \fff^{I_2}_{J_2}$
 is a shorthand of:\smallskip
 
 \noindent
$\bigwedge_{I'_1\subseteq I_1, J'_1\subseteq J_1, I'_2\subseteq I_2, J'_2\subseteq J_2}$\\\vspace{1mm}
$[ \forall x,y,y'\;\;
 {\text \small H}(\spg{I'_1}{J'_1},x,y) \wedge {\text \small H}(\spg{I'_2}{J'_2},x,y')\Rightarrow$\\
\hspace*{25mm}  $\exists x'\;\; {\text \small H}(\spr{I_1\setminus I'_1}{J_1\setminus J'_1},x',y)\wedge {\text \small H}(\spr{I_2\setminus I'_2}{J_2\setminus J'_2},x',y')]$\\
              $\wedge $\\

              $[ \forall x,y,y'\;\;{\text \small H}(\spr{I'_1}{J'_1},x,y) \wedge {\text \small H}(\spr{I'_2}{J'_2},x,y')\Rightarrow  $\\
\hspace*{25mm} $ \exists x'\;\; {\text \small H}(\spg{I_1\setminus I'_1}{J_1\setminus J'_1},x',y)\wedge {\text \small H}(\spg{I_2\setminus I'_2}{J_2\setminus J'_2},x',y')]
               $\\

               \noindent
and $\fff^{I_1}_{J_1}\oaA \fff^{I_2}_{J_2}$ is a shorthand of the formula:\smallskip

 \noindent
$\bigwedge_{I'_1\subseteq I_1, J'_1\subseteq J_1, I'_2\subseteq I_2, J'_2\subseteq J_2}$\\
$[ \forall x,x',y\;\;  {\text \small H}(\spg{I'_1}{J'_1},x,y) \wedge {\text \small H}(\spg{I'_2}{J'_2},x',y)]\Rightarrow$\\
\hspace*{25mm} $\exists y'\; {\text \small H}(\spr{I_1\setminus I'_1}{J_1\setminus J'_1},x,y')\wedge {\text \small H}(\spr{I_2\setminus I'_2}{J_2\setminus J'_2},x',y')]$\\
              $\wedge $ \\
              $[ \forall x,x',y\;\;  {\text \small H}(\spr{I'_1}{J'_1},x,y) \wedge {\text \small H}(\spr{I'_2}{J'_2},x',y)\Rightarrow  $\\
\hspace*{25mm} $ \exists y'\; {\text \small H}(\spg{I_1\setminus I'_1}{J_1\setminus J'_1},x,y')\wedge {\text \small H}(\spg{I_2\setminus I'_2}{J_2\setminus J'_2},x',y')]
$\\
}

Horrible. But this is only because we wrote the rules as  FOL formulas, while they are actually easy to explain in the natural language:
$\fff^{I_1}_{J_1}\obB \fff^{I_2}_{J_2}$ means that whenever two edges can be found in the current swarm, leading from $x$ to $y$ and from $x$ to $y'$,
labelled with two spiders ${\cal S}_1$ and ${\cal S}_2$, of the same color, such that, according to the rule of  Spider Algebra $\clubsuit$, query
$\fff^{I_1}_{J_1}$ can be applied to  ${\cal S}_1$ and $\fff^{I_2}_{J_2}$ can be applied to  ${\cal S}_2$, there must be also a vertex $x'$ in this swarm,
with edges from $x'$ to $y$ and from $x'$ to $y'$,
labelled with  spiders $\fff^{I_1}_{J_1}({\cal S}_1)$ and $\fff^{I_2}_{J_2}({\cal S}_2)$. And analogously for $\fff^{I_1}_{J_1}\oaA \fff^{I_2}_{J_2}$.

So, at Abstraction Level One we no longer need to think about the details of spider anatomy but we are still constrained by the,
 hardly intuitive, rule $\clubsuit$. At Abstraction Level 2 we are going to liberate ourselves also from this constraint.\medskip

\noindent
{\bf Abstraction Level 2 language.} 
Let now ${\mathbb A}_2$ be the subset of  ${\mathbb A}$ consisting of all green (ideal) spiders which are of the form $\spg{I}{}$.
The signature of Abstraction Level 2 language consists of one binary relation  ${\text\small H}({\cal S},\_,\_)$ for each  ${\cal S}\in {\mathbb A}_2$.
 A structure over this signature will be called a {\em green graph}. 
 
 Notice that there is a natural bijection between 
 ${\mathbb A}_2$ and 
 ${\bar{\mathbb S}}
 ={\mathbb S}\cup \{\emptyset \}$, 
 so we will often write\footnote
 {Having two alternative notations for the same object looks like asking for confusion. But see: when discussing
 Precompilation we need to relate Level Two language to Level One, so it is natural to use the notation where spiders are explicit. 
 But then, in Sections \ref{sep-example} and \ref{najwazniejsza} we do not need spiders any more, and we would hate to have the complicated tuples there as small subscripts.
 So $\bar {\mathbb S}$ fits us better than ${\mathbb A}_2$ there.
 }
 $H_i(x,y)$ instead of  
 ${\text\small H}
 (\spg{\{i\}}{},x,y)$ 
 and $H_\emptyset(x,y)$ instead of 
 ${\text\small H}(\spg{}{},x,y)$.

 Concerning the set 
 $\mathbb{L}_2$, of {\em green graph rewriting rules}, for each four spiders  $\spg{I_{1}}{}\neq\spg{I_{3}}{}$, $\spg{I_{2}}{}\neq\spg{I_{4}}{}\in {\mathbb A}_2$,
 there will be two
 rules in the set $\mathbb{L}_2$, denoted as:
 $  I_{1}   \oAA I_{2} \leftrightarrowtriangle I_{3} \oAA I_{4}$ 
 and as $  I_{1}   \oBB I_{2} \leftrightarrowtriangle I_{3} \oBB I_{4}$ 
 where  $  I_{1}   \oAA I_{2} \leftrightarrowtriangle I_{3} \oAA I_{4}$   is a shorthand of:\smallskip

$\forall x,x' \;\;\;\;\;[\exists y \;\; {\text \small H}(\spg{I_{1}}{},x,y)\wedge  {\text \small H}(\spg{I_{2}}{},x',y)]  \Leftrightarrow$\\
\hspace*{15.5mm} $ [\exists y \;\; {\text \small H}(\spg{I_{3}}{},x,y)\wedge  {\text \small H}(\spg{I_{4}}{},x',y)] $\smallskip

\noindent and  $ I_{1}  \oBB I_{2}   \leftrightarrowtriangle I_{3} \oBB I_{4} $  is a shorthand of:\smallskip

$\forall y,y' \;\;\;\;\;[\exists x   \;\; {\text \small H}(\spg{I_{1}}{},x,y)\wedge  {\text \small H}(\spg{I_{2}}{},x,y')] \Leftrightarrow $\\
\hspace*{15.5mm}  $ [\exists x \;\; {\text \small H}(\spg{I_{3}}{},x,y)\wedge  {\text \small H}(\spg{I_{4}}{},x,y')] $

From now on it will be assumed that 
spiders $\spg{3}{}$ and $\spg{4}{}$ (that is -- sets $\{3\}$ and $\{4\}$) do not occur in our sets of green graph rewriting rules.\smallskip

%%% LEVEL2

\noindent
{\bf Compilation and its correctness.} Swarm rewriting rules, as well as green graph rewriting rules, are first order sentences, and each of them is equivalent to a conjunction of tuple generating dependencies.
So, for a set ${\cal T}\subseteq {\mathbb L}_1$ (or ${\cal T}\subseteq {\mathbb L}_2$) and a swarm (resp. green graph) $\mathbb D$ 
the statement ${\mathbb D}\models \cal T$  makes sense and  also the notion of Chase applies to $\cal T$.

\definitionnn{
For a set ${\cal T}\subseteq {\mathbb L}_1$ let  
Compile$({\cal T})=$\\
\hspace*{12mm}$=\{f\oaa f': f\oaA f'\in {\cal T} \}\cup  \{f\obb f': f\obB f'\in {\cal T} \}$.}
%the  subset of ${\cal T}_{{\mathbb F}^2}$ 
%defined by the following procedure:
%\noindent
%\textbullet~ ${\cal Q} := \{f\oaa f': f\oaA f'\in {\cal T} \}\cup  \{f\obb f': f\obB f'\in {\cal T} \}$;\\
%\textbullet~ Compile$({\cal T}):= {\cal T}_{\cal Q}$.}

Which means ``treat each rule from $\cal T$ as a binary query from ${\mathbb F}^2$''.
%, and then Compile$({\cal T})$ is the set of green-red TGDs generated by your queries''.

\definitionnn{
For a set ${\cal T}\subseteq {\mathbb L}_2$ we define Precompile$({\cal T})\subseteq {\mathbb L}_1$ as the result of the following procedure:

\noindent
\textbullet~ Precompile$({\cal T}):=\{\fff^{1}_{1} \oaA \fff^{2}_{2}, \fff^{3}_{1} \oaA \fff^{4}_{2} , \fff^{3} \oaA \fff^{4}_{3}   \} $\\
\textbullet~ fix any numbering of the rules of ${\cal T}$ using  natural numbers  $2,3...k$;\\
\textbullet~ for $i=2$ to $k$,\\
if the $i$'th rule in ${\cal T}$ is 
$  I_{1}   \oAA I_{2} \leftrightarrowtriangle I_{3} \oAA I_{4}$ then add to Precompile$({\cal T})$
the rules
$\fff^{I_{1}}_{2i+1} \oaA \fff^{I_{2}}_{2i+2} $ and $\fff^{I_{3}}_{2i+1} \oaA \fff^{I_{4}}_{2i+2} $\\
and if the $i$'th rule in ${\cal T}$ is 
$ I_{1}   \oBB I_{2} \leftrightarrowtriangle I_{3} \oBB I_{4}$ then add to Precompile$({\cal T})$
the rules 
$\fff^{I_{1}}_{2i+1} \obB \fff^{I_{2}}_{2i+2} $ and $\fff^{I_{3}}_{2i+1} \obB \fff^{I_{4}}_{2i+2} $

}

\begin{remark}\label{inwestycja}
The idea behind  rules of the form $\fff^{I_{1}}_{2i+1} \oaA \fff^{I_{2}}_{2i+2} $ and $\fff^{I_{3}}_{2i+1} \oaA \fff^{I_{4}}_{2i+2} $
in Precompile$({\cal T})$ is that they simulate (in a swarm), in two steps, one execution of 
${I_{1}} \oAA {I_{2}} \leftrightarrowtriangle  {I_{3}} \oAA {I_{4}} $ (in a green graph). As a by-product, two 
red edges are added to the swarm -- labelled with $\spr{}{2i+1}$ and $\spr{}{2i+2}$ 
(the same for $\obB$ instead of $\oaA$).
\end{remark}

\definitionnn{
\begin{itemize}
\item Let ${\cal Q}\subseteq {\mathbb  F^2}$. We will say that  $\cal Q$ {\em leads to the red spider} (or {\em finitely leads to the red spider})
if and only if  each (resp. each finite) structure $\mathbb D$ over $\bar\Sigma$, 
such that ${\mathbb D}\models {\cal T}_{\cal Q}$,  which contains a copy of the full green spider $\spg{}{}$, also contains a copy of the full red spider $\spr{}{}$.

\item Let ${\cal T}\subseteq {\mathbb L}_1$. We will say that  $\cal T$ {\em leads to the red spider} (or {\em finitely leads to the red spider}) if 
each (resp. each finite) swarm $\mathbb D$ such that ${\mathbb D}\models \cal T$,  which contains an atom of the relation ${\text \small H}(\spg{}{},\_,\_)$, also contains 
an atom of the relation ${\text \small H}(\spr{}{},\_,\_)$.

\item We say that a green graph contains a 1-2 pattern if contains  
edges ${\text \small H}(\spg{1}{},a,b)$ and ${\text \small H}(\spg{2}{},a',b)$ for some vertices $a,a',b$.

\item 
Let ${\cal T}\subseteq {\mathbb L}_2$. We will say that  $\cal T$ {\em leads to the red spider} (or {\em finitely leads to the red spider}) if 
each (resp. each finite) green graph $\mathbb D$ such that ${\mathbb D}\models \cal T$  which contains an atom of the relation ${\text \small H}(\spg{}{},\_,\_)$ also contains  
a 1-2 pattern\footnote{
Notice that it would make no sense to repeat,  for ${\cal T}\subseteq {\mathbb L}_2$, the definition  
for ${\cal T}\subseteq {\mathbb L}_1$. This is because a green graph never has any  atom of the relation ${\text \small H}(\spr{}{},\_,\_)$.
But (and please see it as an exercise) having a 1-2 pattern available, the set of rules $Precompile({\cal T})$ will produce, in three steps, using rules 
$\fff^{1}_{1} \oaA \fff^{2}_{2}$, $\fff^{3}_{1} \oaA \fff^{4}_{2}$ , $\fff^{3} \oaA \fff^{4}_{3}$,  
an atom of  ${\text \small H}(\spr{}{},\_,\_)$.
 }.
\end{itemize}
}

\lemmaaa{\label{o-kompilacji}
\begin{enumerate}
\item 
Let ${\cal T}\subseteq {\mathbb L}_1$. Then  $\cal T$ {\em leads to the red spider} (or finitely leads to the red spider) if and only if 
Compile(${\cal T}$) does. 
\item Let ${\cal T}\subseteq {\mathbb L}_2$. Then  $\cal T$ {\em leads to the red spider} (or finitely leads to the red spider) if and only if 
Precompile(${\cal T}$) does. 
\end{enumerate}
}

Proof of the Lemma can be found in Appendix A. 

\noindent
 {\bf It is important to notice} (see condition \ding{183} in Section \ref{g-r}) that:
 
 \observationnn{ A set of queries ${\cal Q}\subseteq {\mathbb F}^2$ leads (finitely leads) to the red spider if and only if it (finitely) determines $\exists^* dalt(\spg{}{})$ (where $\exists^*$ means that all free variables are 
quantified, leading to a boolean query).
 So, by Lemma \ref{o-kompilacji}, since both Precompilation and Compilation are computable, in order to prove Theorem \ref{main1b},
 it suffices to show that it is undecidable for a set  ${\cal T}\subseteq {\mathbb L}_2$ whether $\cal T$ finitely leads to the red spider.
}
 
 From now on proof of Theorem \ref{main1b} has nothing to do with spiders. It is all about green graphs and their rewriting rules.

\section{A separating example}\label{sep-example}

In this Section we are going to prove:

\theoremmm{\label{separacja}
There exists  a set $\sep \subseteq \mathbb{L}_2 $ of 
green graph rewriting rules which does not lead to the red spider, but finitely leads to the red spider.
}

Notice that, by Lemma \ref{o-kompilacji}, this will imply that the set  $Compile(Precompile(\sep))$ of conjunctive queries over $\Sigma$ does not determine the query 
$\exists^*dalt(\spg{}{})$ but finitely determines it.

Here is how we are going to construct $\sep$:

\noindent
{\bf Step 1.} 
Let ${\mathbb D}_{\spg{}{}}$ be a green graph containing just two vertices $\bf a,b$ and one edge $H_\emptyset({\bf a},{\bf b})$ 
(the constants $\bf a$ and $\bf b$ from ${\mathbb D}_{\spg{}{}}$ will be important, please befriend them).
We will define a set $\sep_\infty$ of green graph rewriting rules
 such that $chase(\sep_\infty, {\mathbb D}_{\spg{}{}})$ is an infinite green graph
(a sort of {\em infinite path}) without a 1-2 pattern. 

\noindent
{\bf Step 2.} Another set $\sep_\boxplus$ of green graph rewriting rules will be constructed, 
and $\sep $ will be defined as the union of $\sep_{\infty}$
and  $\sep_\boxplus$. The rules of $\sep_\boxplus$ will be quite complicated (or at least numerous) and it will follow from the construction that
$\sep$ finitely leads to the red spider.

\noindent
{\bf Step 3.}  We will construct (an infinite) green graph $\mathbb M$, containing  ${\mathbb D}_{\spg{}{}}$, without the 1-2 pattern,
and such that  ${\mathbb M}\models \sep$.\\

\noindent
{\bf Step 1.}  Let $\sep_\infty$ consist of three green graph rewriting rules:
(I) $ \emptyset \oAA \emptyset \leftrightarrowtriangle \alpha \oAA \eta_1 $~
(II) $ \emptyset \oBB \eta_1 \leftrightarrowtriangle  \eta_0 \oBB \beta_1 $~
(III) $ \emptyset \oAA \eta_0 \leftrightarrowtriangle \eta_1 \oAA \beta_0 $

\noindent
where $\alpha, \beta_0$ and $\eta_0$ are some even numbers from $\mathbb S$, and $\beta_1$ and $\eta_1$ are odd.\smallskip

Let us try to construct $chase(\sep_\infty, {\mathbb D}_{\spg{}{}})$. We begin from a single 
edge $H_\emptyset({\bf a},{\bf b})$. The only rule that can be applied to this structure is  (I). A new vertex $b_1$ is then created, together 
with two edges: $H_{\alpha}({\bf a},b_1)$ and  $H_{\eta_1}({\bf a},b_1)$. In the next step the only way is to use rule (II),
as we have $H_\emptyset({\bf a},{\bf b})$ and $H_{\eta_1}({\bf a},b_1)$ sharing the beginning vertex. We get a new vertex $a_1$ and new edges
$H_{\eta_0}(a_1,{\bf b})$ and $H_{\beta_1}(a_1,b_1)$. In the third step we can apply rule (III) to 
$H_\emptyset({\bf a},{\bf b})$ and $H_{\eta_0}(a_1,{\bf b})$, creating $b_2$ with
$H_{\eta_1}({\bf a},b_2)$ and $H_{\beta_0}(a_1,b_2)$. And so on -- using rules (II) and (III) alternately we construct 
(see Fig. 1)
 infinite
sequences $b_1$, $b_2\ldots$ of vertices with out-degree 0 and  $a_1$, $a_2\ldots$ of vertices with in-degree 0, connected, in a regular way, with edges labelled with $\beta_0$ and $\beta_1$.\smallskip 

%\begin{figure}
%\label{alphabetapath}
%\input{alpha-beta-path}
%\caption{Finite part of $chase(\sep_\infty, {\mathbb D}_{\spg{}{}})$. }
%\end{figure}

\begin{figure}
\label{alfybety}
%\hspace{-2mm}
\begin{center}
\includegraphics[scale=0.4]{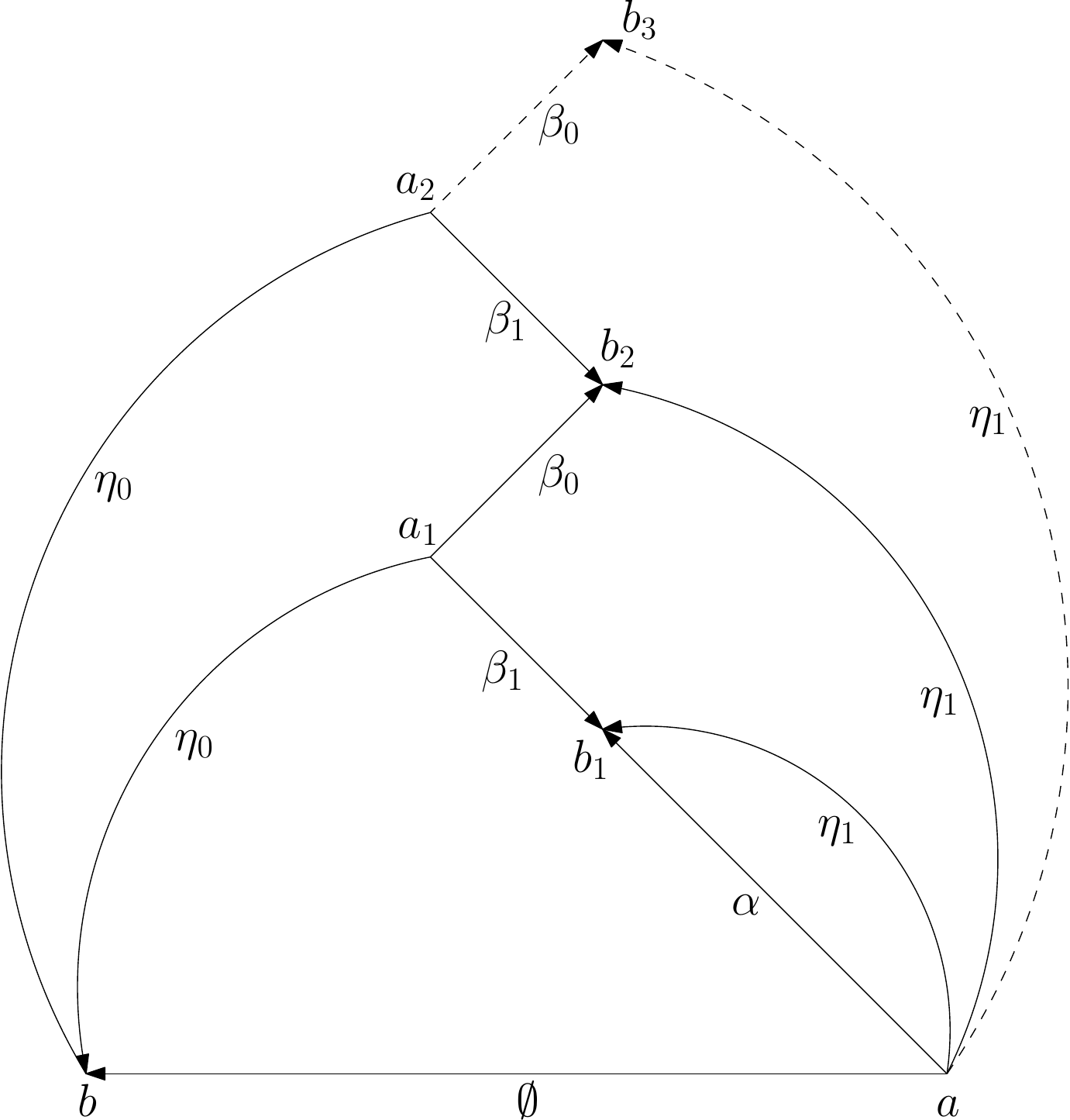}
\end{center}
\vspace{-2.5mm}
\caption{The structure $chase(\sep_\infty, {\mathbb D}_{\spg{}{}})$ {\em in statu nascendi}. Notice (this will be important in Section \ref{outline}) that 
$chase_{i+1}(\sep_\infty, {\mathbb D}_{\spg{}{}})$ is always a result of exactly one application of a rule from $\sep_\infty$ to elements of  
$chase_{i}(\sep_\infty, {\mathbb D}_{\spg{}{}})$.}
\vspace{-2.5mm}
\end{figure}

\noindent
{\bf Seeing a green graph as a set of words (in parity glasses).}
It is standard to see paths in a directed labelled graph as words:

\definitionnn{
Let $\mathfrak M$ be a green graph 
%whose edges are labelled with labels from $\mathbb S$, 
with $s,t$ among its vertices. Then paths$({\mathfrak M},s,t)$ is the set of 
all 
such words ${\mathfrak w}\in {\mathbb S}^*$  
that,
had $\mathfrak M$ been a nondeterministic finite automaton, with the initial state $s$ and a single accepting state $t$,
it would accept ${\mathfrak w}$, but it would not accept any nonempty proper prefix of ${\mathfrak w}$.
}

It will be crucial for us to see green graphs as  complicated sets of words. But unfortunately in the interesting graphs,
all directed paths will have length 1. Each vertex will either have out-degree 0 or in-degree 0 (see Figure 1). This is 
why we need Parity Glasses (reminder: elements of $\mathbb S$ are natural numbers).

\definitionnn{Let $\mathfrak M$ be a green graph containing ${\mathbb D}_{\spg{}{}}$ (which means that $\mathfrak M$ has an edge $H_{\spg{}{}}({\bf a,b})$). Then:\\
\textbullet~ PG($\mathfrak M$) is the graph resulting from $\mathfrak M$ by:
\begin{enumerate}
\item Removing all the edges labelled with $\emptyset$;
\item reversing the direction of all edges labelled with odd numbers.
\end{enumerate}

\noindent\textbullet~$words({\mathfrak M})=\;$paths$(PG({\mathfrak M}),{\bf a,a})\;\cup\; $paths$(PG({\mathfrak M}),{\bf a,b})$.

}

\noindent
{\em Example.} $words(chase(\sep_\infty, {\mathbb D}_{\spg{}{}}))=$\\
\hspace*{10mm}\hfill$=\{\alpha(\beta_1\beta_0)^k\eta_1:k\in{\mathbb N}\}
\cup\{\alpha(\beta_1\beta_0)^k\beta_1\eta_0 :k\in{\mathbb N}\}$\\

By an $\alpha\beta$-path in a green graph $\mathbb M$ we mean a sequence of edges (or vertices, it will be always clear from the context), which 
 seen as a word in $PG({\mathbb M})$ is of the form 
 $\alpha(\beta_1\beta_0)^*$. There are infinitely many $\alpha\beta$-paths in $chase(\sep_\infty, {\mathbb D}_{\spg{}{}})$,
 including for example the path ${\bf a},b_1,a_1,b_2$ and  the path  ${\bf a},b_1,a_1,b_2,a_3,b_3$ (see Figure 1).\smallskip

\noindent
{\bf Step 2.} When a set $\cal T$ of green graph rewriting rules leads to the red spider then this fact is -- at least in principle -- easy to prove. 
One just builds $chase({\cal T}, \mathbb{D}_{\spg{}{}})$ until a 1-2 pattern emerges. But how 
could we possibly prove that $\sep$ (which still remains to be defined, but which is {\bf not} going to lead to the red spider)  does {\bf finitely} lead to the red spider?  

The last means that a 1-2 pattern must emerge in {\em every} finite model of $\cal T$ containing  $\mathbb{D}_{\spg{}{}}$. Is there anything that 
we know for sure about {\em every} finite model of $\cal T$ containing  $\mathbb{D}_{\spg{}{}}$? 
Yes, we know\footnote{The fact that Chase is a universal structure is one of the textbook facts of database theory [JK82].} that for each such model $\mathbb M$
there exists a homomorphism $h:chase({\cal T}, \mathbb{D}_{\spg{}{}})\rightarrow \mathbb M$. Our $\sep$ is going to be a superset of $\sep_\infty$  so
$chase(\sep, \mathbb{D}_{\spg{}{}})$ will contain, as a substructure, the structure $chase(\sep_\infty, \mathbb{D}_{\spg{}{}})$ that we analyzed in Step 1. This means 
that there for sure  will be two vertices $b_t \neq b_{t'}$, such that (remember -- $\mathbb M$ is finite) $h(b_t)= h(b_{t'})$. This means that 
there are two $\alpha\beta$-paths in $\mathbb M$: $h(a),h(b_1),h(a_1)\ldots h(b_t)$ and $h(a),h(b_1),h(a_1)\ldots h(b_{t'})$, of 
{\bf different lengths}, which share the endpoint 
(see Figure 2, do not look at the grid yet). 

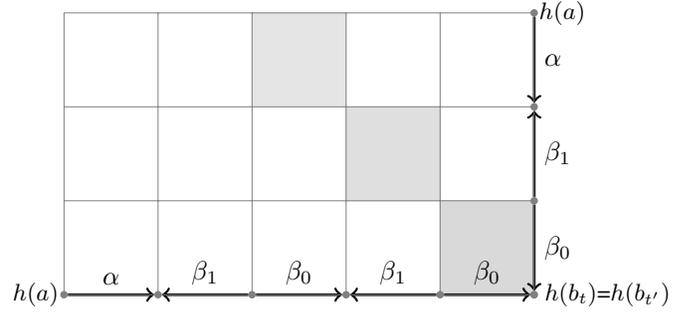
\begin{figure}
\label{p1p2}
\hspace{-2mm}
\begin{tikzpicture}[scale=1.25]
\usetikzlibrary{calc}

\foreach \n in {0} {
\foreach \m in {0,1,2,3,4,5} {
\node[fill=gray, circle, scale=0.3] (n\m\n) at (\m,\n) {$ $};
} 
} 

\filldraw [gray, opacity=0.30] (4,0) rectangle (5,1);
\filldraw [gray, opacity=0.25] (3,1) rectangle (4,2);
\filldraw [gray, opacity=0.20] (2,2) rectangle (3,3);

\foreach \n in {0,1,2,3} {
\foreach \m in {5} {
\node[fill=gray, circle, scale=0.3] (n\m\n) at (\m,\n) {$ $};
} 
}

\path[draw, ->, line width=1pt] (n00) -- (n10) node [pos =0.5, above] {$\alpha$}; 
\path[draw, ->, line width=1pt] (n20) -- (n10) node [pos =0.5, above] {$\beta_1$}; 
\path[draw, ->, line width=1pt] (n20) -- (n30) node [pos =0.5, above] {$\beta_0$}; 
\path[draw, ->, line width=1pt] (n40) -- (n30) node [pos =0.5, above] {$\beta_1$}; 
\path[draw, ->, line width=1pt] (n40) -- (n50) node [pos =0.5, above] {$\beta_0$}; 

\path[draw, ->, line width=1pt] (n53) -- (n52) node [pos =0.5, right] {$\alpha$}; 
\path[draw, ->, line width=1pt] (n51) -- (n52) node [pos =0.5, right] {$\beta_1$}; 
\path[draw, ->, line width=1pt] (n51) -- (n50) node [pos =0.5, right] {$\beta_0$}; 

% \foreach \n in {0.5,1.5,2.5,3.5,4.5} {
% \foreach \m in {0.5,1.5,2.5} {
% \node[fill=gray, square, scale=0.3] (n\m\n) at (\m,\n) {$ $};
% }}

 \node at (-0.3,0) {\small $h(a)$};
\node at (5.3,3)  {\small $h(a)$};
\node at (5.78,0) {\small $h(b_t)$=$h(b_{t'})$};

\draw [help lines] (0,0) grid (5,3);

\end{tikzpicture}
\vspace{-2mm}
\caption{Long initial fragments  of the two paths are also equal in $\mathbb M$ but  we are never going to use this fact, and  it is not indicated on the picture. }
\vspace{-2mm}
\end{figure}

The set $\sep_\boxplus$ will be designed to
detect such two $\alpha\beta$-paths and -- after detecting them -- to create a 1-2 pattern. Figure 2 explains how this will be done: 
the rules of   $\sep_\boxplus$ will build a grid whose
eastern border will be the $\alpha\beta$-path $h({\bf a}),h(b_1),h(a_1)\ldots h(b_t)$ and whose southern border will be the path $h({\bf a}),h(b_1),h(a_1)\ldots h(b_{t'})$. 

This will 
be of course done step by step -- as a result of a single application of a rule from $\sep_\boxplus$ one little square of the grid will be created. For such a step 
a rule will need the southern and eastern edge of the little square to exist, and it will add the western and northern edge\footnote{See -- adding two missing edges of a square is exactly
what green graph rewriting rules are good at.}. After the complete grid is built the rules of $\sep$ will somehow check whether the northwestern corner of the grid
is on the diagonal of the grid. If it is not,
then indeed the $\alpha\beta$-paths  $h({\bf a}),h(b_1),h(a_1)\ldots h(b_t)$ and $h({\bf a}),h(b_1),h(a_1)\ldots h(b_{t'})$ were  of different lengths.

 While the idea is simple, for the real construction we need 41 green graph rewriting rules and 
$4\times2^3=32$ binary relations for the inner edges of the grid (by which we mean edges not belonging to one of the two $\alpha\beta$-paths, 
so that inner vertices, in our sense, also include western and northern border). The names of the 
32 relations (or labels of the green graphs edges) will be conveniently encoded\footnote{
Which, precisely speaking,  means that we assume there is some fixed bijection between our new set of codes and some subset of $\mathbb S$ and that we identify a code with its image under this bijection. We of course assume that this subset is disjoint with $\{\alpha, \beta_0, \beta_1, \eta_0,\eta_1 \}$.
} 
as $\langle n|e|s|w, \alpha|\beta, d|\bar{d}, b|\bar{b}     \rangle$ (where $|$ is the BNF ``or''). 
We think that $\langle n,\alpha,\bar{d},\bar{b}\rangle$ is 1 and $\langle w,\alpha, \bar{d},\bar{b}\rangle$ is 2, where 1 and 2 are the ones from 1-2-pattern.

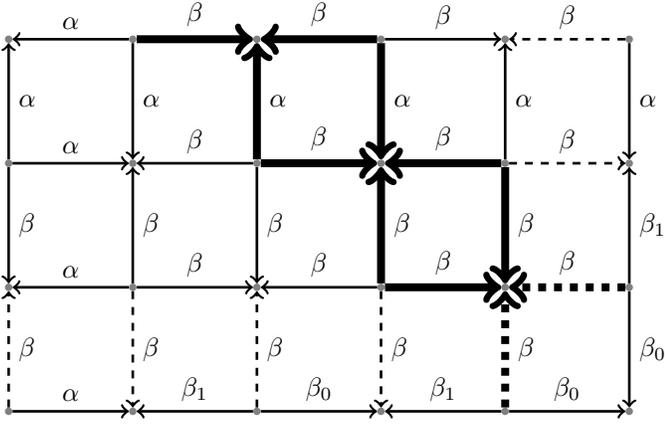
\begin{figure}
\label{c_p1p2}
\begin{tikzpicture}[scale=1.65]
\usetikzlibrary{calc}

\foreach \n in {0,1,2,3} {
\foreach \m in {0,1,2,3,4,5} {
\node[fill=gray, circle, scale=0.3] (n\m\n) at (\m,\n) {$$};
} 
}

\path[draw, ->, line width=1pt] (n00) -- (n10) node [pos =0.5, above] {$\alpha$}; 
\path[draw, ->, line width=1pt] (n20) -- (n10) node [pos =0.5, above] {$\beta_1$}; 
\path[draw, ->, line width=1pt] (n20) -- (n30) node [pos =0.5, above] {$\beta_0$}; 
\path[draw, ->, line width=1pt] (n40) -- (n30) node [pos =0.5, above] {$\beta_1$}; 
\path[draw, ->, line width=1pt] (n40) -- (n50) node [pos =0.5, above] {$\beta_0$}; 

\path[draw, ->, line width=1pt] (n53) -- (n52) node [pos =0.5, right] {$\alpha$}; 
\path[draw, ->, line width=1pt] (n51) -- (n52) node [pos =0.5, right] {$\beta_1$}; 
\path[draw, ->, line width=1pt] (n51) -- (n50) node [pos =0.5, right] {$\beta_0$};

\path[draw,dashed,  ->, line width=1pt] (n00) -- (n01) node [pos =0.5, right] {$\beta$}; 
\path[draw,dashed, ->, line width=1pt] (n11) -- (n10) node [pos =0.5, right] {$\beta$}; 
\path[draw,dashed, ->, line width=1pt] (n20) -- (n21) node [pos =0.5, right] {$\beta$}; 
\path[draw,dashed, ->, line width=1pt] (n31) -- (n30) node [pos =0.5, right] {$\beta$}; 
\path[draw,dashed, ->, line width=3pt] (n40) -- (n41) node [pos =0.5, right] {$\beta$};

\path[draw,  ->, line width=1pt] (n02) -- (n01) node [pos =0.5, right] {$\beta$}; 
\path[draw, ->, line width=1pt] (n11) -- (n12) node [pos =0.5, right] {$\beta$}; 
\path[draw, ->, line width=1pt] (n22) -- (n21) node [pos =0.5, right] {$\beta$}; 
\path[draw, ->, line width=3pt] (n31) -- (n32) node [pos =0.5, right] {$\beta$}; 
\path[draw,->, line width=3pt] (n42) -- (n41) node [pos =0.5, right] {$\beta$}; 

\path[draw,  ->, line width=1pt] (n02) -- (n03) node [pos =0.5, right] {$\alpha$}; 
\path[draw, ->, line width=1pt] (n13) -- (n12) node [pos =0.5, right] {$\alpha$}; 
\path[draw, ->, line width=3pt] (n22) -- (n23) node [pos =0.5, right] {$\alpha$}; 
\path[draw, ->, line width=3pt] (n33) -- (n32) node [pos =0.5, right] {$\alpha$}; 
\path[draw,->, line width=1pt] (n42) -- (n43) node [pos =0.5, right] {$\alpha$};

\path[draw,dashed,  ->, line width=3pt] (n51) -- (n41) node [pos =0.5, above] {$\beta$}; 
\path[draw,dashed, ->, line width=1pt] (n42) -- (n52) node [pos =0.5, above] {$\beta$}; 
\path[draw,dashed, ->, line width=1pt] (n53) -- (n43) node [pos =0.5, above] {$\beta$}; 

\path[draw,  ->, line width=3pt] (n31) -- (n41) node [pos =0.5, above] {$\beta$}; 
\path[draw, ->, line width=3pt] (n42) -- (n32) node [pos =0.5, above] {$\beta$}; 
\path[draw, ->, line width=1pt] (n33) -- (n43) node [pos =0.5, above] {$\beta$}; 

\path[draw,  ->, line width=1pt] (n31) -- (n21) node [pos =0.5, above] {$\beta$}; 
\path[draw, ->, line width=3pt] (n22) -- (n32) node [pos =0.5, above] {$\beta$}; 
\path[draw, ->, line width=3pt] (n33) -- (n23) node [pos =0.5, above] {$\beta$}; 

\path[draw,  ->, line width=1pt] (n11) -- (n21) node [pos =0.5, above] {$\beta$}; 
\path[draw, ->, line width=1pt] (n22) -- (n12) node [pos =0.5, above] {$\beta$}; 
\path[draw, ->, line width=3pt] (n13) -- (n23) node [pos =0.5, above] {$\beta$}; 

\path[draw,  ->, line width=1pt] (n11) -- (n01) node [pos =0.5, above] {$\alpha$}; 
\path[draw, ->, line width=1pt] (n02) -- (n12) node [pos =0.5, above] {$\alpha$}; 
\path[draw, ->, line width=1pt] (n13) -- (n03) node [pos =0.5, above] {$\alpha$};

\end{tikzpicture}
\vspace{-2mm}
\caption{Grid constructed by ${\mathfrak T}_\boxplus$. Diagonal (``$d$'') edges are bold and border (``$b$'') edges are dashed. Labels of the edges in the NW-corner are
$\langle n,\alpha,\bar{d},\bar{b} \rangle$ and $\langle w,\alpha,\bar{d},\bar{b} \rangle$ so they form a 1-2 pattern.}
\vspace{-2mm}
\end{figure}

The first parameter of  a  label of en edge  -- one of $n,e,s,w$ -- 
is the direction the edge heads. The second -- $\alpha$ or $\beta$ -- is inherited from the ``respective'' element
of one of the original $\alpha\beta$-paths. The parameter $d$ (or $\bar{d}$) tells us whether one of the ends of the edge is (or is not) on the diagonal of the grid
(see Figure 3). The fourth parameter (needed in Step 3) tells if an edge shares a vertex with one of the original $\alpha\beta$-paths. 

Now, once we understand the sense of the parameters, it is time to see {\bf the rules of $\sep_\boxplus$}:

$ \beta_0 \oAA \beta_0 \leftrightarrowtriangle  \langle n,\beta,d,b\rangle \oAA \langle w,\beta,d,b\rangle $

The above rule (call it {\em grid triggering rule}) creates the tile in the south-eastern corner of the grid. The next four are:

$ \beta_1 \oBB  \langle n,\beta,d,b\rangle  \leftrightarrowtriangle  \langle s,\beta,\bar{d},b\rangle \oBB \langle e,\beta,d,\bar{b}\rangle $

$ \beta_0 \oAA  \langle s,\beta,\bar{d},b\rangle  \leftrightarrowtriangle  \langle n,\beta,\bar{d},b\rangle \oAA \langle w,\beta,\bar{d},\bar{b}\rangle $

$ \beta_1 \oBB  \langle n,\beta,\bar{d},b\rangle  \leftrightarrowtriangle  \langle s,\beta,\bar{d},b\rangle \oBB \langle e,\beta,\bar{d},\bar{b}\rangle $

$ \alpha \oAA  \langle s,\beta,\bar{d},b\rangle  \leftrightarrowtriangle  \langle n,\beta,\bar{d},b\rangle \oAA \langle w,\alpha,\bar{d},\bar{b}\rangle $

They build the strip of tiles adjacent to the southern edge of the rectangle. Analogously, the strip of tiles adjacent to the eastern edge of the rectangle will be 
built by the four rules:

$ \beta_1 \oBB  \langle w,\beta,d,b\rangle  \leftrightarrowtriangle  \langle e,\beta,\bar{d},b\rangle \oBB \langle s,\beta,d,\bar{b}\rangle $

$ \beta_0 \oAA  \langle e,\beta,\bar{d},b\rangle  \leftrightarrowtriangle  \langle w,\beta,\bar{d},b\rangle \oAA \langle n,\beta,\bar{d},\bar{b}\rangle $

$ \beta_1 \oBB  \langle w,\beta,\bar{d},b\rangle  \leftrightarrowtriangle  \langle e,\beta,\bar{d},b\rangle \oBB \langle s,\beta,\bar{d},\bar{b}\rangle $

$ \alpha \oAA  \langle w,\beta,\bar{d},b\rangle  \leftrightarrowtriangle  \langle w,\beta,\bar{d},b\rangle \oAA \langle n,\alpha,\bar{d},\bar{b}\rangle $

Now last 32 rules of $\sep_\boxplus$ are coming, which will build the interior of the rectangle (including the strips adjacent to the northern and western borders).
Due to the space limit we write them as two schemes, each representing 16 rules:

$ \langle e,\Theta,X,\bar{b}\rangle \oAA  \langle s,\Omega,Y,\bar{b}\rangle  \leftrightarrowtriangle  
  \langle n,\Omega,X,\bar{b}        \rangle \oAA  \langle w,\Theta,Y,\bar{b}\rangle $\\ for each $X,Y\in\{d, \bar{d}\}$ and for each $\Theta, \Omega \in\{\alpha, \beta\}$.

$ \langle w,\Theta,X,\bar{b}\rangle \oBB  \langle n,\Omega,Y,\bar{b}\rangle  \leftrightarrowtriangle  
  \langle s,\Omega,X,\bar{b}        \rangle \oBB  \langle e,\Theta,Y,\bar{b}\rangle $\\ for each $X,Y\in\{d, \bar{d}\}$ and for each $\Theta, \Omega \in\{\alpha, \beta\}$.

%We will refer to the above rules as to {\em the first rule, the 8 rules}, and {\em the 32 rules}.

Now it follows from the construction  that:

\lemmaaa{
$\sep=\sep_\boxplus\cup \sep_\infty  $ finitely leads to the red spider.
}

\noindent
{\bf Step 3.} 
Since now we are going to build an infinite model $\mathbb M$ for  $\sep$,  we can afford having $chase(\sep_\infty, {\mathbb D}_{\spg{}{}})$
as a substructure of $\mathbb M$ and we do not need to identify any of its vertices. But (unfortunately) the grid triggering rule 
still can be applied -- this is since the edges labels
in the left hand side of the first rule of ${\mathfrak T}_\boxplus$ are equal, which means that for each edge $H_{\beta_0}(a_t,b_{t+1})$ in 
$chase(\sep_\infty, {\mathbb D}_{\spg{}{}})$ a new vertex, call it $c_t$ and two new edges, both leading from $a_t$ to $c_t$, will be created, namely
 $ H_{\langle n,\beta, d,b \rangle }(a_t,c_t)$ and $H_{\langle w,\beta,d,b \rangle }(a_t,c_t)$. Then
 the construction from {\bf Step 2} can be repeated, leading to a new grid ${\mathbb M}_t$ (see Figure 4),  constructed
 in the way described in {\bf Step 2} but without a 1-2 pattern. Notice that the picture does not fully reflect the reality here\footnote{ Neither Figure 3 did, but in the 
 context of  {\bf Step 2} we only needed to prove that some pattern {\bf will} appear in the constructed green graph, so -- as we were still able to prove it -- 
 we could pretend that we did not notice that some of the vertices were pairwise equal.} -- in the real green graph the respective vertices of the southern and eastern 
 borders of  ${\mathbb M}_t$ are equal.

Each of ${\mathbb M}_t$ contains some vertices and atoms of $chase(\sep_\infty, {\mathbb D}_{\spg{}{}})$
so, for $t\leq t'$ the intersection of ${\mathbb M}_t$ and ${\mathbb M}_{t'}$ is exactly the $\alpha\beta$-path  ${\bf a},b_1,a_1\ldots b_t$. 
Define ${\mathbb M} = chase(\sep_\infty, {\mathbb D}_{\spg{}{}})\cup\bigcup_{t\in \mathbb N}{\mathbb M}_t$. 
Theorem \ref{separacja} follows from:

\lemmaaa{\label{tugranice}
\begin{enumerate}
\item
${\mathbb M}$ does not contain a 1-2 pattern.
\item
${\mathbb M}\models \sep$
\end{enumerate}
}

For the (easy) proof of Lemma \ref{tugranice} see Appendix B.

 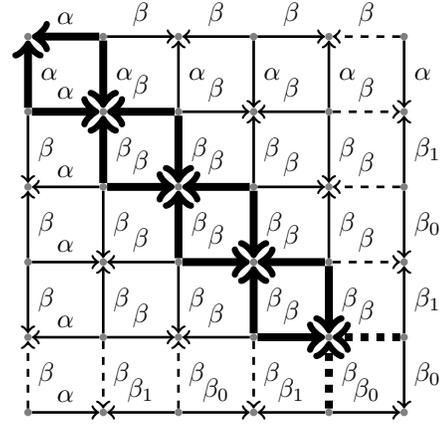
\begin{figure}
 \begin{center}
 \begin{tikzpicture}[scale=1]
\usetikzlibrary{calc}

\foreach \n in {0,1,2,3,4,5} {
\foreach \m in {0,1,2,3,4,5} {
\node[fill=gray, circle, scale=0.3] (n\m\n) at (\m,\n) {$$};
} 
} 

\path[draw, ->, line width=1pt] (n35) -- (n34) node [pos =0.5, right] {$\alpha$}; 
\path[draw, ->, line width=1pt] (n33) -- (n34) node [pos =0.5, right] {$\beta$}; 

\path[draw, ->, line width=3pt] (n15) -- (n14) node [pos =0.5, right] {$\alpha$}; 
\path[draw, ->, line width=3pt] (n13) -- (n14) node [pos =0.5, right] {$\beta$}; 

\path[draw, ->, line width=3pt] (n04) -- (n05) node [pos =0.5, right] {$\alpha$}; 
\path[draw, ->, line width=1pt] (n04) -- (n03) node [pos =0.5, right] {$\beta$};

\path[draw, ->, line width=1pt] (n24)--(n25) node [pos =0.5, right] {$\alpha$}; 
\path[draw, ->, line width=3pt] (n24)--(n23) node [pos =0.5, right] {$\beta$};

\path[draw, ->, line width=1pt] (n44)--(n45) node [pos =0.5, right] {$\alpha$}; 
\path[draw, ->, line width=1pt] (n44)--(n43) node [pos =0.5, right] {$\beta$};

\path[draw, ->, line width=1pt] (n00) -- (n10) node [pos =0.5, above] {$\alpha$}; 
\path[draw, ->, line width=1pt] (n20) -- (n10) node [pos =0.5, above] {$\beta_1$}; 
\path[draw, ->, line width=1pt] (n20) -- (n30) node [pos =0.5, above] {$\beta_0$}; 
\path[draw, ->, line width=1pt] (n40) -- (n30) node [pos =0.5, above] {$\beta_1$}; 
\path[draw, ->, line width=1pt] (n40) -- (n50) node [pos =0.5, above] {$\beta_0$}; 

\path[draw, ->, line width=1pt] (n55) -- (n54) node [pos =0.5, right] {$\alpha$}; 
\path[draw, ->, line width=1pt] (n53) -- (n54) node [pos =0.5, right] {$\beta_1$}; 
\path[draw, ->, line width=1pt] (n53) -- (n52) node [pos =0.5, right] {$\beta_0$}; 
\path[draw, ->, line width=1pt] (n51) -- (n52) node [pos =0.5, right] {$\beta_1$}; 
\path[draw, ->, line width=1pt] (n51) -- (n50) node [pos =0.5, right] {$\beta_0$};

\path[draw,dashed,  ->, line width=1pt] (n00) -- (n01) node [pos =0.5, right] {$\beta$}; 
\path[draw,dashed, ->, line width=1pt] (n11) -- (n10) node [pos =0.5, right] {$\beta$}; 
\path[draw,dashed, ->, line width=1pt] (n20) -- (n21) node [pos =0.5, right] {$\beta$}; 
\path[draw,dashed, ->, line width=1pt] (n31) -- (n30) node [pos =0.5, right] {$\beta$}; 
\path[draw,dashed, ->, line width=3pt] (n40) -- (n41) node [pos =0.5, right] {$\beta$};

\path[draw,  ->, line width=1pt] (n02) -- (n01) node [pos =0.5, right] {$\beta$}; 
\path[draw, ->, line width=1pt] (n11) -- (n12) node [pos =0.5, right] {$\beta$}; 
\path[draw, ->, line width=1pt] (n22) -- (n21) node [pos =0.5, right] {$\beta$}; 
\path[draw, ->, line width=3pt] (n31) -- (n32) node [pos =0.5, right] {$\beta$}; 
\path[draw,->, line width=3pt] (n42) -- (n41) node [pos =0.5, right] {$\beta$}; 

\path[draw,  ->, line width=1pt] (n02) -- (n03) node [pos =0.5, right] {$\beta$}; 
\path[draw, ->, line width=1pt] (n13) -- (n12) node [pos =0.5, right] {$\beta$}; 
\path[draw, ->, line width=3pt] (n22) -- (n23) node [pos =0.5, right] {$\beta$}; 
\path[draw, ->, line width=3pt] (n33) -- (n32) node [pos =0.5, right] {$\beta$}; 
\path[draw,->, line width=1pt] (n42) -- (n43) node [pos =0.5, right] {$\beta$};

\path[draw,dashed,  ->, line width=3pt] (n51) -- (n41) node [pos =0.5, above] {$\beta$}; 
\path[draw,dashed, ->, line width=1pt] (n42) -- (n52) node [pos =0.5, above] {$\beta$}; 
\path[draw,dashed, ->, line width=1pt] (n53) -- (n43) node [pos =0.5, above] {$\beta$}; 
\path[draw,dashed,  ->, line width=1pt] (n44) -- (n54) node [pos =0.5, above] {$\beta$}; 
\path[draw,dashed, ->, line width=1pt] (n55) -- (n45) node [pos =0.5, above] {$\beta$};

\path[draw,  ->, line width=3pt] (n31) -- (n41) node [pos =0.5, above] {$\beta$}; 
\path[draw, ->, line width=3pt] (n42) -- (n32) node [pos =0.5, above] {$\beta$}; 
\path[draw, ->, line width=1pt] (n33) -- (n43) node [pos =0.5, above] {$\beta$}; 
\path[draw, ->, line width=1pt] (n44) -- (n34) node [pos =0.5, above] {$\beta$}; 
\path[draw, ->, line width=1pt] (n35) -- (n45) node [pos =0.5, above] {$\beta$}; 

\path[draw,  ->, line width=1pt] (n31) -- (n21) node [pos =0.5, above] {$\beta$}; 
\path[draw, ->, line width=3pt] (n22) -- (n32) node [pos =0.5, above] {$\beta$}; 
\path[draw, ->, line width=3pt] (n33) -- (n23) node [pos =0.5, above] {$\beta$}; 
\path[draw, ->, line width=1pt] (n24) -- (n34) node [pos =0.5, above] {$\beta$}; 
\path[draw, ->, line width=1pt] (n35) -- (n25) node [pos =0.5, above] {$\beta$};

\path[draw,  ->, line width=1pt] (n11) -- (n21) node [pos =0.5, above] {$\beta$}; 
\path[draw, ->, line width=1pt] (n22) -- (n12) node [pos =0.5, above] {$\beta$}; 
\path[draw, ->, line width=3pt] (n13) -- (n23) node [pos =0.5, above] {$\beta$}; 
\path[draw, ->, line width=3pt] (n24) -- (n14) node [pos =0.5, above] {$\beta$}; 
\path[draw, ->, line width=1pt] (n15) -- (n25) node [pos =0.5, above] {$\beta$}; 

\path[draw,  ->, line width=1pt] (n11) -- (n01) node [pos =0.5, above] {$\alpha$}; 
\path[draw, ->, line width=1pt] (n02) -- (n12) node [pos =0.5, above] {$\alpha$}; 
\path[draw, ->, line width=1pt] (n13) -- (n03) node [pos =0.5, above] {$\alpha$}; 
\path[draw, ->, line width=3pt] (n04) -- (n14) node [pos =0.5, above] {$\alpha$}; 
\path[draw, ->, line width=3pt] (n15) -- (n05) node [pos =0.5, above] {$\alpha$}; 

\end{tikzpicture}
 \end{center}
 \vspace{-2mm}
\caption{Grid ${\mathbb M}_3$. No 1-2 pattern.}
\vspace{-2mm}
 %\label{p3p3}
 \end{figure}

\section{Proof of Theorem \ref{main1b}}\label{najwazniejsza}

\noindent
In $\sep_\infty$ we had $\eta_0$ and $\eta_1$ calling each other in an infinite loop, and creating an infinite 
$\alpha\beta$-path. By homomorphism argument, in presence of such path, the rules of $\sep_\boxplus$   
lead to a 1-2 pattern.
Now this simple mutual recursive call  will be  
controlled by something undecidably complicated -- a rainworm. 

\subsection{\sc Rainworm machine. And how it creeps.}

Rainworm machine (RM), which we  will now define, is a version of (oblivious) Turing Machine, with 
(potentially) right-infinite tape. Alike standard Turing Machine a rainworm machine is described by the finite set of states $\mathfrak{Q}$,
finite tape alphabet $\mathfrak{A}$, finite set of instructions $\Delta$, and an initial configuration.

The ''head'' of a rainworm machine is always located {\bf not} above one of the cells (like in a usual Turing Machine), 
but between two consecutive cells (or right to the rightmost cell). So a configuration of an RM can be in a natural  way seen as 
a word from the language $(\mathfrak{A}+\mathfrak{Q})^*$.

The set of states $\mathfrak{Q}$ of a rainworm machine
is a disjoint union of
${\mathfrak{Q}}^\shortrightarrow_0$,  ${\mathfrak{Q}}^\shortleftarrow_0$, 
${\mathfrak{Q}}^\shortrightarrow_1$,  ${\mathfrak{Q}}^\shortleftarrow_1$, 
of ${\mathfrak{Q}}^\shortrightarrow_{\gamma_0}$,  ${\mathfrak{Q}}^\shortrightarrow_{\gamma_1}$
and of $\{\eta_{11},\eta_0,\eta_1 \}$. 

The finite tape alphabet $\mathfrak{A}$ is a disjoint union of 
sets ${\mathfrak A}_0$, ${\mathfrak A}_1$ and  $\{\alpha, \beta_0, \beta_1, \gamma_0, \gamma_1, \omega_0\}$.

The set $\Delta$ of instructions  consists of 
some number of instructions of any of the following forms\footnote{Do not give up! An informal explanation will soon come.}:\smallskip

\noindent $\diamondsuit_1$:~
 $\eta_{11} \rightsquigarrow \gamma_1\eta_0$\smallskip
 
 \noindent $\diamondsuit_2$:~
 $\eta_0 \rightsquigarrow {\mathfrak b}\eta_1$ where ${\mathfrak b}\in{\mathfrak A}_0 $ \smallskip

\noindent $\diamondsuit_3$:~
$\eta_1 \rightsquigarrow {\mathfrak q}\omega_0$  where ${\mathfrak q}\in \mathfrak{Q}^\shortleftarrow_1$\smallskip

\noindent $\diamondsuit_{4}$:~
${\mathfrak b}'{\mathfrak q} \rightsquigarrow
{\mathfrak q}'{\mathfrak b}$ 
where ${\mathfrak q}\in {\mathfrak{Q}}^\shortleftarrow_0$, ${\mathfrak q}'\in {\mathfrak{Q}}^\shortleftarrow_1$,
 ${\mathfrak b}\in{\mathfrak A}_0$ and ${\mathfrak b}'\in{\mathfrak A}_1 $ 
 
 \noindent $\diamondsuit_{4'}$:~
${\mathfrak b}{\mathfrak q}' \rightsquigarrow
{\mathfrak q}{\mathfrak b}'$ 
where ${\mathfrak q}\in {\mathfrak{Q}}^\shortleftarrow_0$, ${\mathfrak q}'\in {\mathfrak{Q}}^\shortleftarrow_1$,
 ${\mathfrak b}\in{\mathfrak A}_0$ and ${\mathfrak b}'\in{\mathfrak A}_1 $\smallskip
 
 \noindent $\diamondsuit_{5}$:~
 $\gamma_1{\mathfrak q} \rightsquigarrow
\beta_1 {\mathfrak q}'$ 
where ${\mathfrak q}\in {\mathfrak{Q}}^\shortleftarrow_0$, ${\mathfrak q}'\in {\mathfrak{Q}}^\shortrightarrow_{\gamma_0}$

\noindent $\diamondsuit_{5'}$:~
$\gamma_0{\mathfrak q} \rightsquigarrow
\beta_0 {\mathfrak q}'$ 
where ${\mathfrak q}\in {\mathfrak{Q}}^\shortleftarrow_1$, ${\mathfrak q}'\in {\mathfrak{Q}}^\shortrightarrow_{\gamma_1}$\smallskip

 %%%%%%

%%%%%
 
 \noindent $\diamondsuit_6$:~
 ${\mathfrak q}{\mathfrak b} \rightsquigarrow
\gamma_1{\mathfrak q}'$ 
where ${\mathfrak q}\in {\mathfrak{Q}}^\shortrightarrow_{\gamma_1}$, ${\mathfrak q}'\in {\mathfrak{Q}}^\shortrightarrow_0$, ${\mathfrak b}\in{\mathfrak A}_0$

\noindent $\diamondsuit_{6'}$:~
${\mathfrak q}{\mathfrak b} \rightsquigarrow
\gamma_0{\mathfrak q}'$ 
where ${\mathfrak q}\in {\mathfrak{Q}}^\shortrightarrow_{\gamma_0}$, ${\mathfrak q}'\in {\mathfrak{Q}}^\shortrightarrow_1$, ${\mathfrak b}\in{\mathfrak A}_1 $\smallskip

  %%%%%%
 
 \noindent $\diamondsuit_7$:~
 ${\mathfrak q}'{\mathfrak b} \rightsquigarrow
{\mathfrak b}'{\mathfrak q}$ 
where ${\mathfrak q}\in {\mathfrak{Q}}^\shortrightarrow_0$, ${\mathfrak q}'\in {\mathfrak{Q}}^\shortrightarrow_1$,
 ${\mathfrak b}\in{\mathfrak A}_0$ and ${\mathfrak b}'\in{\mathfrak A}_1 $

 \noindent $\diamondsuit_{7'}$:~
 ${\mathfrak q}{\mathfrak b}' \rightsquigarrow 
{\mathfrak b}{\mathfrak q}'$ 
 where ${\mathfrak q}\in {\mathfrak{Q}}^\shortrightarrow_0$, ${\mathfrak q}'\in {\mathfrak{Q}}^\shortrightarrow_1$,
 ${\mathfrak b}\in{\mathfrak A}_0$ and ${\mathfrak b}'\in{\mathfrak A}_1 $\smallskip
 
%%%%%%

 \noindent $\diamondsuit_8$:~
$  {\mathfrak q}\omega_0 \rightsquigarrow {\mathfrak b}\eta_0$  where ${\mathfrak q}\in \mathfrak{Q}^\shortrightarrow_1$ and ${\mathfrak b}\in{\mathfrak A}_1 $ \smallskip
 
 We require  the set $\Delta$ of instructions to be a partial
 function\footnote{In other words, this condition means that rainworm machine is a deterministic computation model.} -- two different instructions must  have different left hand sides.
 
 The initial configuration of the machine is $\alpha\eta_{11}$.
 
As we said before, a configuration of an RM can be in a natural  way seen as 
a word from the language $(\mathfrak{A}+\mathfrak{Q})^*$, so $\Delta$ is formulated in the language of 
 Thue semisystem rules\footnote{See e.g. our paper [GM15]. Or [D77] if you prefer a more serious introduction}. 
 A single computation step of an RM can (should) be seen as a 
single application of a Thue semi-system rewriting. Following the standard Thue systems notational  convention the notation
${\mathfrak w}\rightsquigarrow_\vartriangle {\mathfrak v}$ means that ${\mathfrak w}={\mathfrak w}_1{\mathfrak s}{\mathfrak w}_2$ and
 ${\mathfrak v}={\mathfrak w}_1{\mathfrak t}{\mathfrak w}_2$ for some rule ${\mathfrak s}\rightsquigarrow{\mathfrak t}$ in  $\Delta$.
 We also use $\stackrel{k\;\;}{\rightsquigarrow_\vartriangle}$ to denote the $k$'th power of relation  $\rightsquigarrow_\vartriangle$,   
 $\stackrel{\ast\;\;}{\rightsquigarrow_\vartriangle}$ to denote the transitive closure of $\rightsquigarrow_\vartriangle$ and 
 $\stackrel{\ast\;\;}{ \leftrightsquigarrow_\vartriangle}$ to denote the symmetric transitive closure of $\rightsquigarrow_\vartriangle$ (i.e. the smallest equivalence 
 relation having $\rightsquigarrow_\vartriangle$ as a subset).
 
While a configuration of a rainworm machine can be always seen as a word, it is clear that not all words from  $(\mathfrak{A}+\mathfrak{Q})^*$ make sense as configurations:

\definitionnn{\label{rm-konfiguracja}
Call symbols in  $\{\alpha, \beta_0, \gamma_0, \eta_0\}\cup {\mathfrak{Q}}^\shortrightarrow_0 \cup {\mathfrak{Q}}^\shortleftarrow_0\cup {\mathfrak{Q}}^\shortrightarrow_{\gamma_0}\cup {\mathfrak{A}}_0$ 
{\em even} and symbols in  $\{\beta_1, \gamma_1, \eta_1, \eta_{11}\}\cup {\mathfrak{Q}}^\shortrightarrow_1 \cup {\mathfrak{Q}}^\shortleftarrow_1\cup {\mathfrak{Q}}^\shortrightarrow_{\gamma_1}\cup {\mathfrak{A}}_1$ {\em odd}. 
A word ${\mathfrak w}\in(\mathfrak{A}+\mathfrak{Q})^*$ is an RM configuration, if:
 \begin{enumerate}

\item 
${\mathfrak w}\in\mathfrak{A}^+\mathfrak{Q}\mathfrak{A}^*$ (which means there is exactly one symbol in  ${\mathfrak w}$ which symbolizes the head of the machine);

\item the last symbol of $\mathfrak w$ is one of $\eta_{11}$, $\eta_0$,  $\eta_1$, $\omega_0$;

\item odd and even symbols occur in
${\mathfrak w}$ alternately (there are never two odd or two even symbols next to each other);
 \item
 ${\mathfrak w}$ is of the form ${\mathfrak w}_1{\mathfrak w}_2$ where ${\mathfrak w}_1$ is of the form $\alpha(\beta_1\beta_0)^*$ or 
 $\alpha(\beta_1\beta_0)^*\beta_1$, 
 ${\mathfrak w}_2$ begins with $\gamma_0$ or $\gamma_1$ or an element of ${\mathfrak{Q}}^\shortrightarrow_{\gamma_0}$
 or an element of ${\mathfrak{Q}}^\shortrightarrow_{\gamma_1}$,
 and none of $\alpha$, $\beta_0$, $\beta_1$ occur in ${\mathfrak w}_2$.

\end{enumerate}

}

Proof of the following lemma is straightforward case inspection and induction:

\lemmaaa{\label{oczywistyRM} 
Let $\alpha\eta_{11} \stackrel{\ast\;\;}{\rightsquigarrow_\vartriangle} \mathfrak w$. Then ${\mathfrak w}$ is an RM configuration.
}

Now we are going to explain -- informally -- how a rainworm creeps. Imagine ${\mathfrak w}$  like in Lemma \ref{oczywistyRM}
and let $={\mathfrak w}_1{\mathfrak w}_2$ be as in Definition \ref{rm-konfiguracja}(4). Suppose
the last  symbol of ${\mathfrak w}_2$ is $\eta_0$. Think of ${\mathfrak w}_2$ as of a rainworm ($\eta_0$ being its front and $\gamma_0$ or $\gamma_1$ being its rear end) 
and of ${\mathfrak w}_1$ as of the slime trail a rainworm
leaves behind. Now there is at most one thing we can do (since $\Delta$ is a  partial function) -- we can use some rule of the form
$\diamondsuit_2$ to rewrite $\eta_0$ into ${\mathfrak b}\eta_1$ for some ${\mathfrak b}$. Our rainworm has grown one symbol longer! 
In the next step we can\footnote{We should repeat here -- and in several other places in this paragraph -- that ``there is at most one thing we can do now''.} use a rule of the form $\diamondsuit_3$ 
to rewrite $\eta_1$ into ${\mathfrak q}\omega_0$ for some ${\mathfrak q}\in \mathfrak{Q}^\shortleftarrow_1$. The rainworm has grown one symbol longer again, 
but now the head\footnote{Head in the Turing machine sense, rainworms have neither head nor tail. They have front and rear.} is no longer in front 
of the animal. The state of the head is from $\mathfrak{Q}^\shortleftarrow_1\cup \mathfrak{Q}^\shortleftarrow_0$ now.
The head will now move, cell by cell, towards the rear end of the rainworm 
(applying rules $\diamondsuit_4$ and $\diamondsuit_4'$ alternately) rewriting the symbols from ${\mathfrak A}_0 \cup {\mathfrak A}_1$ it passes on its way. 
Then, after $\gamma_1$ (or $\gamma_0$) is reached, it is rewritten, by some rule of he form  $\diamondsuit_5$ (or $\diamondsuit_5'$) into $\beta_1$ (or $\beta_0$).
The rear end of the rainworm moves towards the front,  but the slime trail gets one symbol longer!

Now -- since the last rule used was one of $\diamondsuit_5$ (or $\diamondsuit_5'$) -- the state is one from   
${\mathfrak{Q}}^\shortrightarrow_{\gamma_0}$ (or  ${\mathfrak{Q}}^\shortrightarrow_{\gamma_1}$),
so the next rewriting will move the head to the right and replace the first symbol from ${\mathfrak A}_1$ it encounters
with $\gamma_0$ (or, resp. the first symbol from ${\mathfrak A}_0$ with $\gamma_1$). Only rules of the form $\diamondsuit_7$ or $\diamondsuit'_7$
will be applicable then, and the head will keep moving right (and rewriting the tape symbols on its way), towards the front of the rainworm, until $\omega_0$ is found.
Then a rule of the form $\diamondsuit_8$ will be used, and we will be back to the original situation, with $\eta_0$ as the rightmost symbol. 
Notice that the rainworm is longer now (we added one symbol twice and removed one symbol once) and also the slime trail (which is an $\alpha\beta$-path) is longer (we added one symbol). Given $\Delta$, there are of course two possibilities -- either the rainworm will creep forever, leaving behind an infinite  $\alpha\beta$-slime trail or, 
at some point,  no rule will be applicable, and the process will terminate. It is easy to prove, using textbook techniques, that:

\lemmaaa{\label{z-dzdzownica-nie-wiadomo}The problem whether, for given $\Delta$,  the rainworm\footnote{Given $\Delta$, the sets $\mathfrak Q$ and $\mathfrak A$ can be reconstructed.} creeps forever, is undecidable.}

\subsection{\sc Creeping back and forth}

It is not going to surprise anyone that our next goal is to translate $\Delta$ into a set of green graph rewriting rules. 
 But, unlike rainworm machine instructions, green graph rewriting rules are symmetric. This is how we deal with it:
 
 \lemmaaa{\label{jakis-porzadek}
 \begin{enumerate}
 \item
 If  ${\mathfrak w}\stackrel{\ast\;\;}{ \rightsquigarrow_\vartriangle} {\mathfrak v}$ and ${\mathfrak v}$ is an RM-configuration then
${\mathfrak w}$ satisfies conditions (1)-(3) of Definition \ref{rm-konfiguracja}. 
 \item
 If  ${\mathfrak w}$ satisfies condition (1) of Definition \ref{rm-konfiguracja} then there exists at most one  ${\mathfrak v}$ such that 
  ${\mathfrak w}\rightsquigarrow_\vartriangle {\mathfrak v}$. 
  \item
  There is a constant $c_\vartriangle$, such that if 
  ${\mathfrak v}$ satisfies condition (1) of Definition \ref{rm-konfiguracja} then there exist at most $c_\vartriangle$ words  ${\mathfrak w}$ such that 
  ${\mathfrak w}\rightsquigarrow_\vartriangle {\mathfrak v}$.
% \item
% If  ${\mathfrak w}\stackrel{\ast\;\;}{ \rightsquigarrow_\vartriangle} {\mathfrak v}$ then ${\mathfrak w}$ is at most as long as ${\mathfrak v}$.
 \end{enumerate}
 }
 
 \noindent
{\em Proof:} First claim can be proved by straightforward case inspection and induction. 
The second and the third  follow easily from the construction of $\Delta$ (remember that $\Delta$ is a partial function).\eop
 
 {\bf Now suppose}, till the end of this subsection,  {\bf that the computation of a  rainworm machine with the set of instructions $\Delta$ terminates}, after some 
 number $k_\vartriangle$ of steps, which means that there is ${\mathfrak u}_\vartriangle$ 
 such that  $\alpha\eta_{11}\stackrel{\;k_\vartriangle\;}{ \rightsquigarrow_\vartriangle} {\mathfrak u}_\vartriangle$ and no rule from $\Delta $ can be applied to ${\mathfrak u}_\vartriangle$ any more. Then:

 \lemmaaa{ \label{jakis-porzadek-2}
 \begin{enumerate}
 \item
 $\{{\mathfrak w}: {\mathfrak w}\stackrel{\ast\;\;}{\leftrightsquigarrow_\vartriangle} \alpha\eta_{11}\}=
 \{{\mathfrak w}:{\mathfrak w}\stackrel{\ast\;\;}{ \rightsquigarrow_\vartriangle} {\mathfrak u}_\vartriangle\}$.

 \item
 If 
 ${\mathfrak w}\stackrel{\ast\;\;}{ \rightsquigarrow_\vartriangle} {\mathfrak u}_\vartriangle$ then $\mathfrak w$ satisfies cond. (4)
 of Definition \ref{rm-konfiguracja}.
 \item If
 ${\mathfrak w}\stackrel{k\;}{ \rightsquigarrow_\vartriangle} {\mathfrak u}_\vartriangle$ then $k\leq k_\vartriangle$.

 \item The set  $\{{\mathfrak w}:{\mathfrak w}\stackrel{\ast\;\;}{ \rightsquigarrow_\vartriangle} {\mathfrak u}_\vartriangle\}$ is finite.

 \end{enumerate}
 }

 \noindent
{\em Proof:} First claim follows from Lemma \ref{jakis-porzadek}: to reach any vertex of a tree (and,  due to Lemma \ref{jakis-porzadek}(2) 
 the interesting part of the $\rightsquigarrow_\vartriangle$-graph is a tree) from a leaf, it is enough to go up to the root and then down. 
  For the proof of the second claim notice that it of course holds for ${\mathfrak u}_{\tiny \Delta}$,
 and that if it holds for some ${\mathfrak v}_2$, and ${\mathfrak v}_1\rightsquigarrow_\vartriangle {\mathfrak v}_2$ then it also holds for 
 ${\mathfrak v}_1$. 
 
 To see why the third claim is true notice that it follows from the construction of $\Delta$ 
 that, while of course for a configuration $\mathfrak w$ of $\Delta$ and for 
 a $k\in \mathbb N$ there may very well be two different configurations  $\mathfrak v$ and  $\mathfrak v'$ such that 
 ${\mathfrak v}\stackrel{k\;}{ \rightsquigarrow_\vartriangle} {\mathfrak w}$ and  ${\mathfrak v'}\stackrel{k\;}{ \rightsquigarrow_\vartriangle} {\mathfrak w}$,
 always in such case ${\mathfrak v}$ and ${\mathfrak v}'$ will be of equal length and 
 the machine head (that is the symbol from $\mathfrak Q$)  will be in the same place in   ${\mathfrak v}$ and ${\mathfrak v}'$.
 
 \noindent
 Last claim follows from the third and from Lemma \ref{jakis-porzadek}(3).  \eop

\subsection{\sc From rainworms to green graph rules.}

For a rainworm machine $\Delta$ we define the set $\sep_\vartriangle$ of green graph rewriting rules as follows:\smallskip

\noindent
\textbullet~ Rules    $\emptyset\oAA\emptyset  \leftrightarrowtriangle \alpha\oAA \eta_{11}$     
and
 $\eta_{11}\oBB \emptyset \leftrightarrowtriangle   \gamma_1\oBB \eta_0$ are
  in $\sep_\vartriangle$;\smallskip
 
\noindent
\textbullet~
$\eta_0 \oAA \emptyset \leftrightarrowtriangle {\mathfrak b}\oAA \eta_1$ is in  $\sep_\vartriangle$ if
 $\eta_0 \rightsquigarrow {\mathfrak b}\eta_1$ is in  $\Delta$;\smallskip
 
\noindent
\textbullet~
$\eta_1 \oBB \emptyset \leftrightarrowtriangle {\mathfrak q}\oBB \omega_0$ is in  $\sep_\vartriangle$ 
if $\eta_1 \rightsquigarrow  {\mathfrak q}\omega_0$ is in  $\Delta$;\smallskip

\noindent
\textbullet~
$ {\mathfrak x} \oBB {\mathfrak t} \leftrightarrowtriangle  {\mathfrak x}' \oBB {\mathfrak t}'$ is in  $\sep_\vartriangle$ 
if $ {\mathfrak x}{\mathfrak t}   \rightsquigarrow  {\mathfrak x}'{\mathfrak t}'$ is an instruction of the form 
$\diamondsuit_4$, $\diamondsuit_5$, $\diamondsuit_6$,  $\diamondsuit_7$ or  $\diamondsuit_8$ in  $\Delta$;\smallskip

\noindent
\textbullet~
$ {\mathfrak x} \oAA {\mathfrak t} \leftrightarrowtriangle  {\mathfrak x}' \oAA {\mathfrak t}'$ is in  $\sep_\vartriangle$ 
if $ {\mathfrak x}{\mathfrak t}   \rightsquigarrow  {\mathfrak x}'{\mathfrak t}'$ is an instruction of the form 
$\diamondsuit_4'$, $\diamondsuit_5'$, $\diamondsuit_6'$ or $\diamondsuit_7'$ in  $\Delta$.\smallskip

Now, {\bf in view of Lemma \ref{o-kompilacji} and Lemma \ref{z-dzdzownica-nie-wiadomo}, in order to prove Theorem \ref{main1b} it is enough to show:}

\lemmaaa{\label{rownowazne-z-dzdz}  For given $\Delta$,  the rainworm creeps forever (i.e. rainworm machine with $\Delta$ as its set of instructions never halts)
if and only if the set ${\mathfrak T}^\boxplus_\vartriangle= {\mathfrak T}_\vartriangle\cup {\mathfrak T}_\boxplus $ of green graph rewriting rules finitely leads to the red spider.
}

\noindent
The rest of this Section is devoted to the proof of Lemma \ref{rownowazne-z-dzdz}

\subsection{\sc The {\em ``$\Rightarrow$'' } direction (the easier one).}
Suppose $\Delta$ is the set of instructions of some rainworm which creeps forever. Then:

\lemmaaa{\label{krolik}
~\\
If
 $\alpha\eta_{11} \stackrel{\ast\;\;}{\rightsquigarrow_\vartriangle} \mathfrak w$ then ${\mathfrak w}\in
words(chase(\sep_\vartriangle, {\mathbb D}_{\spg{}{}}))$.
}

\noindent
{\em Proof:} Of course $\alpha\eta_{11}\in
words(chase(\sep_\vartriangle, {\mathbb D}_{\spg{}{}}))$, since one gets it from ${\mathbb D}_{\spg{}{}}$ by a single application of the first rule of  $\sep_\vartriangle$.

\noindent
For the induction step we are going to show that if  ${\mathfrak w}\rightsquigarrow_\vartriangle{\mathfrak v}$  and ${\mathfrak w}\in
words(chase(\sep_\vartriangle, {\mathbb D}_{\spg{}{}}))$ then also ${\mathfrak v}\in
words(chase(\sep_\vartriangle, {\mathbb D}_{\spg{}{}}))$. 

So suppose ${\mathfrak w}={\mathfrak w}_1{\mathfrak s}{\mathfrak w}_2$ and
 ${\mathfrak v}={\mathfrak w}_1{\mathfrak t}{\mathfrak w}_2$ for a rule ${\mathfrak s}\rightsquigarrow{\mathfrak t}$ in  $\Delta$.
 Let (for example, as all cases are similar)  ${\mathfrak s}={\mathfrak c}_0{\mathfrak c}_1$ and ${\mathfrak t}={\mathfrak c'}_0{\mathfrak c'}_1$, for some 
 even ${\mathfrak c}_0$ and    ${\mathfrak c'}_0$ and odd ${\mathfrak c}_1$ and    ${\mathfrak c'}_1$. Then, by construction of 
 $\sep_\vartriangle$,
 $ {\mathfrak c}_0\oAA {\mathfrak c}_1 \leftrightarrowtriangle  {\mathfrak c'}_0\oAA {\mathfrak c'}_1 $ is  a rule of  $\sep_\vartriangle$.

 Let $c,c'$ be vertices of the green graph $chase(\sep_\vartriangle, {\mathbb D}_{\spg{}{}})$ such that:\hfill
 ${\mathfrak w}_1\in paths(PG(chase(\sep_\vartriangle, {\mathbb D}_{\spg{}{}})),{\bf a},c)$,\\
 that \hfill
 ${\mathfrak s}\in paths(PG(chase(\sep_\vartriangle, {\mathbb D}_{\spg{}{}})),c,c')$\\
 and  that \hfill ${\mathfrak w}_2\in paths(PG(chase(\sep_\vartriangle, {\mathbb D}_{\spg{}{}})),c',{\bf a})$\\
(or \hfill ${\mathfrak w}_2\in paths(PG(chase(\sep_\vartriangle, {\mathbb D}_{\spg{}{}})),c',{\bf b})$).

Now, ${\mathfrak s}\in paths(PG(chase(\sep_\vartriangle, {\mathbb D}_{\spg{}{}})),c,c')$ 
means that there is a vertex $d$ of $chase(\sep_\vartriangle, {\mathbb D}_{\spg{}{}})$ such that 
$H_{{\mathfrak c}_0}(c,d)$ and $H_{{\mathfrak c}_1}(c',d)$ are edges of $chase(\sep_\vartriangle, {\mathbb D}_{\spg{}{}})$.
 But, since
$chase(\sep_\vartriangle, {\mathbb D}_{\spg{}{}})\models\sep_\vartriangle$, the rule 
$ {\mathfrak c}_0\oAA {\mathfrak c}_1 \leftrightarrowtriangle  {\mathfrak c'}_0\oAA {\mathfrak c'}_1 $ is also satisfied in $chase(\sep_\vartriangle, {\mathbb D}_{\spg{}{}})$,
and thus there exists a vertex $d'$ in $chase(\sep_\vartriangle, {\mathbb D}_{\spg{}{}})$ such that 
$H_{{\mathfrak c'}_0}(c,d')$ and $H_{{\mathfrak c'}_1}(c',d')$ are edges of $chase(\sep_\vartriangle, {\mathbb D}_{\spg{}{}})$.
This means that ${\mathfrak t}\in paths(PG(chase(\sep_\vartriangle, {\mathbb D}_{\spg{}{}}),c,c')$ and, in consequence, ${\mathfrak v}\in
words(chase(\sep_\vartriangle, {\mathbb D}_{\spg{}{}}))$. 
\eop

Since rainworm $\Delta$ creeps forever, it follows from Lemma \ref{krolik} that there are $\alpha\beta$-paths of unbounded length in 
$chase(\sep_\vartriangle, {\mathbb D}_{\spg{}{}})$, and hence in $chase(\sep^\boxplus_\vartriangle, {\mathbb D}_{\spg{}{}})$. Now we use the machinery
from Section \ref{sep-example} Step 2 to prove that every finite model of $\sep^\boxplus_\vartriangle$, containing $ {\mathbb D}_{\spg{}{}} $, contains a 1-2 pattern, so 
 $\sep^\boxplus_\vartriangle$ finitely leads to the red spider.

%Like $chase(\sep_\infty, {\mathbb D}_{\spg{}{}})$, also  $chase(\sep_\vartriangle, {\mathbb D}_{\spg{}{}})$ contains 

\subsection{\sc The {\em ``$\Leftarrow$'' } direction (the harder one).}

Now we consider a rainworm $\Delta$ whose computation terminates. Let ${\mathfrak u}_\vartriangle=\alpha(\beta_1\beta_0)^n\gamma_1\ldots \omega_0 $ 
be the final configuration\footnote{This is only to fix attention and simplify notations. Of course 
 ${\mathfrak u}_\vartriangle$ does not need to be exactly of this form -- the last symbol can also be $\eta_0$, $\eta_1$ 
 or even $\eta_{11}$, and the symbol after last $\beta_0$ may very well be different than $\gamma_1$.} like in Section \ref{najwazniejsza} B, and let also 
 $k_\vartriangle$ be as defined there. 
Our goal is to construct a finite green graph $\mathfrak M$, without a 1-2 pattern, containing ${\mathbb D}_{\spg{}{}}$ 
and being a model of $\sep^\boxplus_\vartriangle$. It would be tempting to think that $chase(\sep^\boxplus_\vartriangle, {\mathbb D}_{\spg{}{}})$ is such a 
graph. But for some subtle reasons, it is not. Apparently, already  $chase(\sep_\vartriangle, {\mathbb D}_{\spg{}{}})$ can be infinite.

First we will apply a chase-like procedure to construct a finite structure $\mathbb M$ being a model of $\sep_\vartriangle$.

Let ${\mathbb M}^0$ be the structure containing just nothing more than ${\mathbb D}_{\spg{}{}}$ and ${\mathfrak u}_\vartriangle$: 
there is an edge $H_\emptyset({\bf a},{\bf b})$, 
then  edges $H_\alpha({\bf a},b_1)$, $H_{\beta_1}(a_1,b_1)$,  $H_{\beta_0}(a_1,b_2)\ldots H_{\beta_1}(a_n,b_n)$ and so on, and finally there 
is an edge, labelled with $\omega_0$, from some vertex to $\bf b$. The structure
${\mathbb M}$ is defined by the following {\bf procedure}:\smallskip

%\noindent\textbullet~ ${\mathbb M}:={\mathbb M}_0$;

\noindent
\textbullet~ {\bf for} m=0 {\bf to} $k_\vartriangle$ {\bf do}:\\
$\{$
${\mathbb M}^{m+1}:={\mathbb M}^m$;

\noindent
{\bf for all} 
pairs  $c,c'$ of vertices of ${\mathbb M}^m$  
and {\bf all} rules of the form
$ {\mathfrak c}\oAA {\mathfrak d} \leftrightarrowtriangle  {\mathfrak c'}\oAA {\mathfrak d'} $ 
[or of the form  $ {\mathfrak c}\oBB {\mathfrak d}\leftrightarrowtriangle  {\mathfrak c'}\oBB {\mathfrak d'} $] in $\sep_\vartriangle$
{\bf if:}\\
there exists vertex $d'$ of ${\mathbb M}^m$ such that:\\
\mbox{(\hspace{-0.4mm}$\spadesuit$\hspace{-0.4mm}) ${\mathbb M}^m \hspace{-1.5mm}\models H_{{\mathfrak c'}}(c,d'), H_{{\mathfrak d'}}(c'\hspace{-0.7mm},d')$ [or ${\mathbb M}^m\hspace{-1.5mm}\models H_{{\mathfrak c'}}(d'\hspace{-0.7mm},c), H_{\mathfrak d'}(d'\hspace{-0.7mm},c')$]}\\
 {\bf but:}\\
($\heartsuit$)~ there is no such vertex $d$ of ${\mathbb M}^m$  that:\\
 ${\mathbb M}^m\models H_{{\mathfrak c}}(c,d), H_{{\mathfrak d}}(c',d)$ [or  ${\mathbb M}^m\models H_{{\mathfrak c}}(d,c), H_{{\mathfrak d}}(d,c')$]\\\smallskip
 \noindent
{\bf do:}$\{$\\
{\bf (i)} if ${\mathfrak d}\neq \emptyset $, then add to $\mathbb M^{m+1}$ a new vertex $d$ and edges  $H_{{\mathfrak c}}(c,d)$ and $H_{{\mathfrak d}}(c',d)$ 
[resp. edges $H_{{\mathfrak c}}(d,c)$, $H_{{\mathfrak d}}(d,c')$];\smallskip

\noindent
{\bf (ii)} if ${\mathfrak d} = \emptyset $, then  add to $\mathbb M^{m+1}$ an edge  $H_{{\mathfrak c}}(c,{\bf b})$ 
[resp. an edge $H_{{\mathfrak c}}({\bf a},c)$ ~];\\
$\}\}$;\\
\textbullet~  ${\mathbb M}:={\mathbb M}^{k_\vartriangle+1}$.\medskip

For three reasons this procedure does not ({\em a priori}) build $chase(\sep_\vartriangle, {\mathbb D}_{\spg{}{}})$. First,
while each rule in $\sep_\vartriangle$ is an equivalence (a conjunction of two TGDs), the procedure executes only one of them, the right-to-left one. 
Second, in ({\bf ii}), instead of (as {\em chase} would do) creating a vertex $d$ and two edges $H_{\mathfrak c}(c,d)$ and $H_\emptyset({\bf a},d)$ it 
reuses\footnote{
It is easy to prove that in the situation of ({\bf ii}) we have $c'={\bf a}$ [or $c'={\bf b}$].}
the existing vertex $\bf b$ and edge $H_\emptyset({\bf a, b})$. Third reason is that it terminates after a fixed number of stages, possibly before
reaching a fixpoint.

 \lemmaaa{\label{bezsenny}
 \begin{itemize}
\item
 ${\mathbb M}$ is finite and ${\mathbb M}\models \sep_\vartriangle$.
 \item
 If ${\mathbb M}\models H_{\beta_0}(x,y)$, 
 (or  ${\mathbb M}\models H_{\beta_1}(x,y)$) for some vertices $x,y$, then $x$, $y$ are 
 already vertices of ${\mathbb M}^0$ and ${\mathbb M}^0\models H_{\beta_0}(x,y)$, (or  ${\mathbb M}^0\models H_{\beta_1}(x,y)$).
 \end{itemize}
 }

Proof of Lemma \ref{bezsenny} can be found in Appendix C. 

Now consider ${\mathfrak M}=chase({\mathfrak T}^\boxplus, {\mathbb M})$. No two different edges labelled with $\beta_0$ share the end in ${\mathbb M}$. But, as we saw
in Section \ref{sep-example} (Step 3) this does not stop the grid triggering rule from being applied. But again, exactly like in 
 Section \ref{sep-example} (Step 3)
for each $t\leq n$ (where $n$ is the one 
from ${\mathfrak u}_\vartriangle$ above) nothing more than a harmless grid ${\mathbb M}_t$ will be constructed, not having 
1-2-pattern\footnote{
Notice that ${\mathbb M}_t$ has, among others, elements  $a_i$ and $b_i$ for all $i\leq t$, and $\mathbb M$ also has, among others, 
elements  $a_i$ and $b_i$ for all $i\leq t$. This is not a coincidence -- the set union that defines ${\mathfrak M}$ would make no sense otherwise. }. 
So ${\mathfrak M}={\mathbb M}\cup \bigcup_{t\leq n}{\mathbb M}_t$. Since the only edges in ${\mathfrak M}$ which were not in $\mathbb M$ are 
grid edges, we still have ${\mathfrak M}\models \sep_\vartriangle$, so ${\mathfrak M}$ is indeed a finite model of $\sep^\boxplus_\vartriangle$, without the 1-2 pattern.

%%%%%%%%%%%%%%%%%%%%%%%%%%%%%%%%%%%%%%%%%%%%%%%%%%%%%%%%%%%%%%%%%%%%%%%REWR%%%%%%%%%%%%%%%%%%%%%%%%%%%%%%%%%%%%%%%%%%%%%%%%%%%%%%%%%%%%%%%%%%%%%%  

%%%%%%%%%%%%%%%%%%%%%%%%%%%%%%%%%%%%%%%%%%%%%%%%%%%%%%%%%%%%%%%%%%%%%%%REWR%%%%%%%%%%%%%%%%%%%%%%%%%%%%%%%%%%%%%%%%%%%%%%%%%%%%%%%%%%%%%%%%%%%%%%  
%%%%%%%%%%%%%%%%%%%%%%%%%%%%%%%%%%%%%%%%%%%%%%%%%%%%%%%%%%%%%%%%%%%%%%%%%%%%%%%%%%%%%%%%%%%%%%%%%%%%%%%%%%%%%%%%%%%%%%%%%%%%%%%%%%%%%%%%%%%%%
\section{FO non-rewritability. Proof of Theorem \ref{nieprzepisywalnosc} (outline).}\label{outline}

It is not hard to guess what is going to be our  $\cal Q$ and $Q$ (notations as in Section \ref{malutka}) 
-- we only know one example of a set of CQs which finitely determines another query but
does not determine it. So -- let ${\cal Q}=Compile(Precompile(\sep))$ and $Q=\exists^* dalt(\spg{}{})$. 
We need  to produce, for a given  $l\in \mathbb N$, two structures, ${\mathbb D}_y$ and  ${\mathbb D}_n$, over $\Sigma$
 (we are back to Abstraction Level 0) such that  ${\mathbb D}_y$ contains a copy of $dalt(\spg{}{})$, ${\mathbb D}_n$ does not contain one,
 but the views ${\cal Q}({\mathbb D}_y)$ and ${\cal Q}({\mathbb D}_n)$ are not distinguishable by a FO Ehrenfeucht-Fraisse game with $l$ rounds.

Before we go any further we need to  understand well enough the sense of 
${\cal T}_{\cal Q}$ -- the set of green-red TGDs generated by $\cal Q$.
For simplicity, instead of $\cal Q$,  let us concentrate on ${\cal Q}_\infty=Compile(Precompile(\sep_\infty))$. It contains six\footnote{
Plus the three queries $\fff^{1}_{1} \oaa \fff^{2}_{2}, \fff^{3}_{1} \oaa \fff^{4}_{2} , \fff^{3} \oaa \fff^{4}_{3} $, but we do not need to think about them now. 
}
queries:

(IA) $ \fff^{}_{5} \oaa \fff^{}_{6} $   \hfill (IB) $ \fff^{\alpha}_{5} \oaa \fff^{\eta_1}_{6} $  

(IIA) $ \fff^{}_{7} \obb \fff^{\eta_1}_{8}$   \hfill (IIB) $ \fff^{\eta_0}_{7} \obb \fff^{\beta_1}_{8}$  

(IIIA)  $ \fff^{}_{9} \oaa \fff^{\eta_0}_{10}$   \hfill (IIIB)  $ \fff^{\eta_1}_{9} \oaa \fff^{\beta_0}_{10}$ 

 We imagine two viewers -- Grace and Ruby -- each of them is shown her own
structure over $\Sigma$. Grace will see ${\mathbb D}_y$ and Ruby ${\mathbb D}_n$.   The structures are uncolored, but we imagine 
the one 
seen by Grace is Green and the one seen by Ruby is Red. Our goal, as we have already said,  is to have ${\mathbb D}_y$ with $\spg{}{} $,
 ${\mathbb D}_n$  without $\spr{}{}$, but to make sure that each of the 
the two girls can see (almost) the same thing.
The trick is %- of course -- 
that they do not see the real structures, but their image under the CQs in ${\cal Q}_\infty$.

\subsection{Attempt 1 --  ${\mathbb D}_y$ and ${\mathbb D}_n$ as fragments of $chase({\cal T}_{{\cal Q}_\infty},\spg{}{})$.}\label{1attempt}

As we said, there must be a copy of the full green spider $\spg{}{}$ in ${\mathbb D}_y$. Let ${\bf a}$ and ${\bf b}$ be its tail and antenna.
Now, suppose  some current versions of ${\mathbb D}_y$ and ${\mathbb D}_n$ are  defined. One of the viewers, say Grace, can complain 
that ``Ruby has a tuple $\bf t$ in $Q({\mathbb D}_n)$ (for some $Q\in {\cal Q}_\infty$) which I do not see in  $Q({\mathbb D}_y)$''. And, as we want the two girls to see same, we must
add $\bf t$ to $Q({\mathbb D}_y)$. But of course we cannot simply add a tuple to $Q({\mathbb D}_y)$ -- we add something to ${\mathbb D}_y$ in order to
make sure that $\bf t$ is in $Q({\mathbb D}_y)$. And this is exactly what $Q^{R\rightarrow G}$ is about. 

So, as we see $chase$ (as a procedure), with respect to ${\cal T}_{{\cal Q}_\infty}$, is all about making sure that the girls see the same. 
If some tuple $t$ is in $ {{\cal Q}_\infty}(dalt(chase_i({\cal T}_{{{\cal Q}_\infty}},\spg{}{})\rrestriction G))$ then it is also in
$ {{\cal Q}_\infty}(dalt(chase_{i+1}({\cal T}_{{\cal Q}_\infty},\spg{}{})\rrestriction R))$, and the other way round. 
And so, {\bf once 
$chase$ reaches a fixpoint, both girls really  see the same}. But -- and it is an important point  -- the 
structures they watch are very much different. Grace watches the daltonisation of
$chase({\cal T}_{{\cal Q}_\infty},\spg{}{})\rrestriction G $, and -- if we abstract from the $\Sigma$-details and see the structure
as a swarm -- $chase({\cal T}_{{\cal Q}_\infty},\spg{}{})\rrestriction G $
contains edges labelled with $\spg{}{}, \spg{\alpha}{}, \spg{\eta_0}{}, \spg{\eta_1}{}, \spg{\beta_0}{}, \spg{\beta_1}{}$.
Ruby watches the daltonisation of
$chase({\cal T}_{{\cal Q}_\infty},\spg{}{})\rrestriction R $, and -- again seen as a swarm -- $chase({\cal T}_{{\cal Q}_\infty},\spg{}{})\rrestriction R $
contains edges labelled with $\spg{}{4}, \spg{}{5}, \spg{}{6}, \spg{}{7}, \spg{}{8}, \spg{}{9}$. 

But  $chase({\cal T}_{{\cal Q}_\infty},\spg{}{})$ is infinite and we need ${\mathbb D}_y$ and ${\mathbb D}_n$ to be finite. On the other hand,
we do not really require ${{\cal Q}_\infty}({\mathbb D}_y)$ and ${{\cal Q}_\infty}({\mathbb D}_g)$ to be equal. We just need them to be 
similar enough 
 so that $l$-rounds Ehrenfeucht-Fraisse game cannot spot the difference. So how about having, as ${\mathbb D}_y$ and ${\mathbb D}_n$, 
daltonisations of
$chase_i({\cal T}_{{\cal Q}_\infty},\spg{}{})\rrestriction G$ and  of $chase_i({\cal T}_{{\cal Q}_\infty},\spg{}{})\rrestriction R$? We know 
from the analysis of Figure 1 and from Remark \ref{inwestycja} that at each  stage of $chase$ a single execution of a rule from ${{\cal Q}_\infty}$ will be 
performed: $chase$  begins with an application of rule (IA) (or, strictly speaking, with an 
application of a TGD generated by 
%query 
(IA), something red  will be added then), then 
(IB), and then rules (IIA), (IIB), (IIIA), (IIIB) are used alternately, building a complicated green-red structure which, after some abstraction, looks like a 
long green $\alpha\beta$-path (with additional $\eta_0$ and $\eta_1$ edges leading to $\bf a$ and $\bf b$) wrapped with some red edges. 

Since there is always only one match in $chase_i({\cal T}_{{\cal Q}_\infty},\spg{}{})\rrestriction G$ of a TGD generated by one of  the queries in ${{\cal Q}_\infty}$,
the structures ${{\cal Q}_\infty}(dalt(chase_{i}({\cal T}_{{\cal Q}_\infty},\spg{}{})\rrestriction R))$ and 
${{\cal Q}_\infty}(dalt(chase_{i}({\cal T}_{{\cal Q}_\infty},\spg{}{})\rrestriction G))$ will always differ by just one atom. 
Two long ``paths''  which differ by just one atom sounds -- from the point of view of Ehrenfeucht-Fraisse games -- like hope. 
But unfortunately, it follows from the construction of ${{\cal Q}_\infty}$, that the relation Ruby will see watching 
$chase_i({\cal T}_{{\cal Q}_\infty},\spg{}{})\rrestriction R$
via query (IIA) is always equal to the relation she will see watching the same structure via (IIB), and that the same holds for queries (IIIA) and (IIIB).
But the two equalities never hold simultaneously for relations Grace sees in  $chase_i({\cal T}_{{\cal Q}_\infty},\spg{}{})\rrestriction G$. So whatever way we try to 
prematurely terminate this infinite  chase, the single  atom of difference will be enough for a FO formula to spot a difference.

\subsection{Attempt 2 -- ${\mathbb D}_y$ and ${\mathbb D}_n$ for ${{\cal Q}_\infty}$ }\label{attempt2}

Define $chase_{2i}^L({\cal T}_{{\cal Q}_\infty},\spg{}{})$ ($L$ standing for ``late'')  as the set of atoms added to  
$chase_{2i}({\cal T}_{{\cal Q}_\infty},\spg{}{})$
at some stage $j$, where $i\leq j\leq 2i$, together with all elements involved with these atoms (including $\bf a$ and $\bf b$). 
In other words, atoms of $chase_{2i}^L({\cal T}_{{\cal Q}_\infty},\spg{}{})$ are atoms of $chase_{2i}({\cal T}_{{\cal Q}_\infty},\spg{}{})$ which are 
not in $chase_{i}({\cal T}_{{\cal Q}_\infty},\spg{}{})$.

As before, structures
${\cal Q}_\infty(dalt(chase_{2i}^L({\cal T}_{{\cal Q}_\infty},\spg{}{})\rrestriction G))$ and  ${\cal Q}_\infty(dalt(chase_{2i}^L({\cal T}_{{\cal Q}_\infty},\spg{}{})\rrestriction R))$ 
differ, by one atom, at the end of the path. And -- unlike  ${\cal Q}_\infty(dalt(chase_{i}({\cal T}_{{\cal Q}_\infty},\spg{}{})\rrestriction R))$ 
and  ${\cal Q}_\infty(dalt(chase_{i}({\cal T}_{{\cal Q}_\infty},\spg{}{})\rrestriction G))$, they also differ at the ``beginning'': 
their beginnings  are exactly as different as the  ends of ${{\cal Q}_\infty}(dalt(chase_{i}({\cal T}_{{\cal Q}_\infty},\spg{}{})\rrestriction R))$
and of ${{\cal Q}_\infty}(dalt(chase_{i}({\cal T}_{{\cal Q}_\infty},\spg{}{})\rrestriction G))$
are.

Now take $i$ Large Enough (with respect to $l$) and:

\textbullet~ define ${\mathbb D}_y$ as a disjoint union\footnote{Elements $\bf a$ and $\bf b$ are seen as constants,
so the word ``disjoint'' does not apply to them -- they belong to all the copies.}
of $ dalt(chase_{i}({\cal T}_{{\cal Q}_\infty},\spg{}{})\rrestriction G)) $; of $i$ copies of $dalt(chase_{2i}^L({\cal T}_{{\cal Q}_\infty},\spg{}{})\rrestriction G))$
and of $i$ copies of $dalt(chase_{2i}^L({\cal T}_{{\cal Q}_\infty},\spg{}{})\rrestriction R))$;

\textbullet~ define ${\mathbb D}_n$ as a disjoint union
of $ dalt(chase_{i}({\cal T}_{{\cal Q}_\infty},\spg{}{})\rrestriction R)) $; of $i$ copies of $dalt(chase_{2i}^L({\cal T}_{{\cal Q}_\infty},\spg{}{})\rrestriction G))$
and of $i$ copies of $dalt(chase_{2i}^L({\cal T}_{{\cal Q}_\infty},\spg{}{})\rrestriction R))$;

Now an Ehrenfeucht-Fraisse games argument comes, which could not be more standard: 
each girl can see -- via the queries of ${{\cal Q}_\infty}$ -- some number of paths ($2i+1$ of them, to be precise). 
Among the paths Grace can see there is one whose beginning looks the same as the beginning of the Ruby's counterpart of this path
 (it is $ dalt(chase_{i}({\cal T}_{{\cal Q}_\infty},\spg{}{})\rrestriction G)) $), $i$ paths whose beginnings indicate that they come from 
daltonisation of the  green fragment of  $chase_{2i}^L({\cal T}_{{\cal Q}_\infty},\spg{}{})$ and $i$ paths whose beginnings indicate that they come from 
daltonisation of the red fragment of  $chase_{2i}^L({\cal T}_{{\cal Q}_\infty},\spg{}{})$. Notice that  exactly the same collection of left ends is seen 
by Ruby. 

Now the path ends: Grace can see $i+1$  ends which indicate that they come from 
daltonisation of the  green fragment of  $chase_{i}({\cal T}_{{\cal Q}_\infty},\spg{}{})$  or   of  $chase_{2i}^L({\cal T}_{{\cal Q}_\infty},\spg{}{})$ 
(the ends of the two kinds paths 
are 
identical) and $i$  ends which indicate that hey come from 
daltonisation of the  red fragment of  $chase_{i}({\cal T}_{{\cal Q}_\infty},\spg{}{})$  or   of  $chase_{2i}^L({\cal T}_{{\cal Q}_\infty},\spg{}{})$. Since $i$ is Large,
the difference between $i$ and $i+1$ is of course not FO-noticeable\footnote{Saying ``FO'' we  mean FO formula equivalent to some $l$-round game.}. 

Among the paths Grace can see there is one  whose beginning looks the same as the beginning of the Ruby's counterpart of this path
 and whose 
 end indicates being green. The end of the respective path Ruby can see indicates being red. But 
 since the paths are Long (because $i$ is Large) their beginnings cannot be related to their  ends using 
 a FO formula, and in consequence ${\mathbb D}_y$ and ${\mathbb D}_n$ are indeed FO-indistinguishable.
 
 Clearly, there are details that we omitted. Most notably, for Ruby to see anything, 
  via queries (IIA) and (IIIA), in her daltonised copies of 
 $dalt(chase_{i}^L({\cal T}_{{\cal Q}_\infty},\spg{}{})\rrestriction R))$,
 two spiders, $dalt(\spr{}{7})$ and $dalt(\spr{}{9})$, both having $\bf a$ as the tail and $\bf b$ as the antenna
 need to be present in ${\mathbb D}_n$. Also, while this outline is stated 
 mainly on the Abstraction Level 1, the real construction must be at Level 0. 
 
 \subsection{Last step. Structures ${\mathbb D}_y$ and ${\mathbb D}_n$ for ${\cal Q}$. }
 This was already complicated for ${\cal Q}_\infty$, and the construction we really need is for $\cal Q$. Let 
 ${\cal Q}_\boxplus = Compile(Precompile(\sep_\boxplus))$.
 
 Let again  $i$ be a Large Enough natural number and
imagine the structure 
$chase({\cal T}_{{\cal Q}_\boxplus},chase_i({\cal T}_{{\cal Q}_\infty}, \spg{}{}))$ 
(or $chase({\cal T}_{{\cal Q}_\boxplus},chase^L_{2i}({\cal T}_{{\cal Q}_\infty}, \spg{}{}))$ ). Like in the case of structures ${\mathbb M}_t$ in Section \ref{sep-example}, a grid will be
 constructed, with $chase_{i}({\cal T}_{{\cal Q}_\infty},\spg{}{})$ playing the $\alpha\beta$-path being the southern and eastern border of this grid. 
 But now of course the grid is two-colored: we have our old  green grid, intertwining with some red edges, as explained in Remark \ref{inwestycja}.
 The structure of this green-red grid is unclear, but the good news is that we do not need to care: since 
 $chase({\cal T}_{{\cal Q}_\boxplus},chase_i({\cal T}_{{\cal Q}_\infty}, \spg{}{}))$ reaches its fixpoint after a finite number of stages (i.e. terminates), 
we can be sure (as we observed in Section \ref{1attempt}) that  
${\cal Q}_\boxplus(dalt(chase({\cal T}_{{\cal Q}_\boxplus},chase_i({\cal T}_{{\cal Q}_\infty}, \spg{}{}))\rrestriction G))$ and
 ${\cal Q}_\boxplus(dalt(chase({\cal T}_{{\cal Q}_\boxplus},chase_i({\cal T}_{{\cal Q}_\infty}, \spg{}{}))\rrestriction R))$ are simply equal -- 
using the queries in ${\cal Q}_\boxplus$ both girls see the same structures there. So
the difference between 
${\cal Q}(dalt(chase({\cal T}_{{\cal Q}_\boxplus},chase_i({\cal T}_{{\cal Q}_\infty}, \spg{}{}))\rrestriction G))$ and 
${\cal Q}(dalt(chase({\cal T}_{{\cal Q}_\boxplus},chase_i({\cal T}_{{\cal Q}_\infty}, \spg{}{}))\rrestriction R))$ remains to be one atom -- the 
same one which made ${\cal Q}(dalt(chase_{i}({\cal T}_{\cal Q},\spg{}{})\rrestriction R))$ and 
${\cal Q}(dalt(chase_{i}({\cal T}_{\cal Q},\spg{}{})\rrestriction G))$ different (and the difference between 
${\cal Q}(dalt(chase({\cal T}_{{\cal Q}_\boxplus},chase^L_{2i}({\cal T}_{{\cal Q}_\infty}, \spg{}{}))\rrestriction G))$ and 
${\cal Q}(dalt(chase({\cal T}_{{\cal Q}_\boxplus},chase^L_{2i}({\cal T}_{{\cal Q}_\infty}, \spg{}{}))\rrestriction R))$ remains to be two  atoms).

Now define ${\mathbb D}_y$ and ${\mathbb D}_n$ like in Section \ref{attempt2} but instead of each component of the form $ dalt({\cal D}\rrestriction G)$ or 
$ dalt({\cal D}\rrestriction R)$ of any of the two disjoint unions there (which means -- instead of a part of a path) take 
$ dalt(chase({\cal T}_{{\cal Q}_\boxplus}, {\cal D})\rrestriction G)$ and $ dalt(chase({\cal T}_{{\cal Q}_\boxplus}, {\cal D})\rrestriction R)$
(which means -- take the grid having this part of a path as two of its borders). 

Now notice that the crucial argument we needed for FO-indistinguishability in Section \ref{attempt2} was that 
the  ends of the paths in each of ${\mathbb D}_y$ and ${\mathbb D}_n$ are far enough from each other, so that they cannot be related to each other by FO.
The last observation needed to 
see that (the new) ${\mathbb D}_y$ and ${\mathbb D}_n$ will be FO-indistinguishable is that adding the grid to the path does not decrease the distance between 
the ends.

\section{References}

{\small

\noindent
[A11] 
Foto N. Afrati;
  {\em Determinacy and Query Rewriting for Conjunctive Queries and Views},
 Theor. Comput. Sci. 412 vol 11;
 2011;
 pages 1005--1021;\smallskip

\noindent
[AD98] Serge Abiteboul, Oliver M. Duschka;
  {\em Complexity of Answering Queries Using Materialized Views},
Proceedings of 17th ACM SIGACT-SIGMOD-SIGART PODS; 
 1998; pages 254--263;\smallskip

 \noindent
[BGO10]
Vince Barany, Georg Gottlob and Martin Otto,
{\em Querying the guarded fragment};  Proceedings of ACM/IEEE LICS 2010; pages 1–10;\smallskip

\noindent
 [CGLV00] Diego Calvanese, Giuseppe De Giacomo, Maurizio Lenzerini, and Moshe Y Vardi,
{\em Answering regular path queries using views};
Proc. of 16th International Conference on Data Engineering, IEEE, 2000;
 pages 389–398;\smallskip

\noindent
[CR97]
 Chandra Chekuri,
               Anand Rajaraman,
   {\em Conjunctive Query Containment Revisited},
 Proceedings of  ICDT '97;
 Springer, Lecture Notes in Computer Science,
  vol. 1186; 1997; pages 56-70;\smallskip

 \noindent
[D77]
Martin Davis,
{\em Unsolvable Problems},
in {em Handbook of Mathematical Logic},
J. Barwise ed.,
North-Holland 1977;
pages 567-594;\smallskip

\noindent
[DPT99]
Alin Deutsch,
                Lucian Popa and
                Val Tannen,
%   editor    = {Malcolm P. Atkinson and
%                Maria E. Orlowska and
%                Patrick Valduriez and
%                Stanley B. Zdonik and
%                Michael L. Brodie},
{\em Physical Data Independence, Constraints, and Optimization with
            Universal Plans},
Proceedings of  VLDB'99; 
Morgan Kaufmann;  pages 459-470; 1999;\smallskip
%   isbn      = {1-55860-615-7},
%  
%   ee        = {db/conf/vldb/DeutschPT99.html},
%   crossref  = {DBLP:conf/vldb/99},
%   bibsource = {DBLP, http://dblp.uni-trier.de}
% }
% 

\noindent
 [F15] Nadime Francis, PhD Thesis, 2015;\smallskip

\noindent
[FG12]
Enrico Franconi,
               Paolo Guagliardo,
{\em The View Update Problem Revisited},
CoRR,
abs/1211.3016, 2012;\smallskip

%   year      = {2012},
%   url       = {http://arxiv.org/abs/1211.3016},
%   timestamp = {Sat, 01 Dec 2012 20:32:37 +0100},
%   biburl    = {http://dblp.uni-trier.de/rec/bib/journals/corr/abs-1211-3016},
%   bibsource = {dblp computer science bibliography, http://dblp.org}
% }
% 

\noindent
[FGZ12]
 Wenfei Fan, Floris Geerts and Zheng Lixiao,
{\em View Determinacy for Preserving Selected Information in Data Transformations};
Inf. Syst. 37.1; Elsevier 2012; pages 1--12;\smallskip
%  issue_date = {March, 2012},
%  volume = {37},
%  number = {1},
%  month = mar,
%  year = {2012},
%  issn = {0306-4379},
%  pages = {1--12},
%  numpages = {12},
%  url = {http://dx.doi.org/10.1016/j.is.2011.09.001},
%  doi = {10.1016/j.is.2011.09.001},
%  acmid = {2043744},
%  publisher = {Elsevier Science Ltd.},
%  address = {Oxford, UK, UK},
%  keywords = {Information preservation, Queries, Rewriting, View determinacy, Views},
% } 
% 
% 

\noindent
[FKN13] Enrico Franconi,  Volha Kerhet and Ngo Nhung,
{\em Exact Query Reformulation with First-Order Ontologies and Databases}
Journal of Artificial Intelligence Research 48 (2013);\smallskip

\noindent
[FSF14] Nadime Francis,  Luc Segoufin, Cristina  Sirangelo,
{\em Datalog Rewritings of Regular Path Queries using Views},
Proceedings of 17th International Conference on Database Theory (ICDT), Athens, Greece, March 24-28, 2014.,
107--118;
2014;\smallskip

\noindent
[GM13] Tomasz Gogacz, Jerzy Marcinkowski, {\em Converging to the Chase--A Tool for Finite Controllability};
Proceedings of  ACM/IEEE Symposium on LICS; 2013;\smallskip

\noindent
[GM15] Tomasz Gogacz, Jerzy Marcinkowski {\em The Hunt for a Red Spider: Conjunctive Query Determinacy Is Undecidable }; 
ACM/IEEE Symposium on Logic in Computer Science, LICS 2015; pages 281-292;\smallskip

\noindent
[H01]
Alon Y. Halevy, {\em Answering Queries Using Views: A Survey},
The VLDB Journal vol 10.4;\smallskip
%  issue_date = {December 2001},
%  volume = {10},
%  number = {4},
%  month = dec,
%  year = {2001},
%  issn = {1066-8888},
 pages 270--294;  2001; \smallskip
%  numpages = {25},
%  url = {http://dx.doi.org/10.1007/s007780100054},
%  doi = {10.1007/s007780100054},
%  acmid = {767151},
%  publisher = {Springer-Verlag New York, Inc.},
%  address = {Secaucus, NJ, USA},
%  keywords = {Data integration, Date warehousing, Materialized views, Query optimization, Survey, Web-site management},
% } 
% 
% 
% @inproceedings{LMS95,
%  author = {Levy, Alon Y. and Mendelzon, Alberto O. and Sagiv, Yehoshua},
%  title = {Answering Queries Using Views (Extended Abstract)},
%  booktitle = {Proceedings of the Fourteenth ACM SIGACT-SIGMOD-SIGART Symposium on Principles of Database Systems},
%  series = {PODS '95},
%  year = {1995},
%  isbn = {0-89791-730-8},
%  location = {San Jose, California, USA},
%  pages = {95--104},
%  numpages = {10},
%  url = {http://doi.acm.org/10.1145/212433.220198},
%  doi = {10.1145/212433.220198},
%  acmid = {220198},
%  publisher = {ACM},
%  address = {New York, NY, USA},
% } 
% 
% 
% 

\noindent
[JK82]
D.S.Johnson, D. S. and A. Klug,
{\em Testing Containment of Conjunctive Queries Under Functional and Inclusion Dependencies},
 Proceedings of the 1st ACM SIGACT-SIGMOD Symposium on Principles of Database Systems PODS 82,
pages 164--169;\smallskip

\noindent
[NSV07] Alan Nash, Luc Segoufin  and Victor Vianu;
{\em Determinacy and Rewriting of Conjunctive Queries Using Views: A Progress Report},
ICDT 2007, Springer LNCS 4353; pages 59-73;\smallskip

\noindent
[NSV10] Alan Nash, Luc Segoufin  and Victor Vianu;
{\em Views and queries: Determinacy and rewriting},
ACM Trans. Database Syst 35; 2010; pages 21:1{\textendash}21:41\smallskip
% 	month = {July},
% 	pages = {},
% 	publisher = {ACM},
% 	address = {New York, NY, USA},
% 	keywords = {Queries, rewritingg, views},
% 	issn = {0362-5915},
% 	doi = {http://doi.acm.org/10.1145/1806907.1806913},
% 	url = {http://doi.acm.org/10.1145/1806907.1806913},
% 	author = {Nash, Alan and Segoufin, Luc and Vianu, Victor}
% }
% 
% 

\noindent
[P11] Daniel Pasail{\u{a}}, {\em Conjunctive queries determinacy and rewriting}; Proc. ICDT 2011; ACM Press; pages 220-231;\smallskip
%   	address =	{Uppsala, Sweden},
%   	author =	{, Daniel},
%   	booktitle =	{{P}roceedings of the 14th {I}nternational {C}onference on {D}atabase {T}heory ({ICDT}'11)},
%   	DOI =	{10.1145/1938551.1938580},
%   	editor =	{Milo, Tova},
%   	month =	mar,
%   	pages =	{},
%   	publisher =	{},
%   	title =	
%   	url =	{http://www.lsv.ens-cachan.fr/Publis/PAPERS/PDF/pasaila-icdt11.pdf},
%   	year = {2011}
%   	
%   	}
%   	

\noindent
[R06]
Ricardo Rosati,
{\em On the decidability and finite controllability of query processing in
  databases with incomplete information};
Proceedings of ACM PODS 2006;
  pages 356--365.\smallskip

%  @inproceedings{LY85,
%  author = {Larson, Per-Ake and Yang, H. Z.},
%  title = {Computing Queries from Derived Relations},
%  booktitle = {Proceedings of the 11th International Conference on Very Large Data Bases - Volume 11},
%  series = {VLDB '85},
%  year = {1985},
%  location = {Stockholm, Sweden},
%  pages = {259--269},
%  numpages = {11},
%  url = {http://dl.acm.org/citation.cfm?id=1286760.1286784},
%  acmid = {1286784},
%  publisher = {VLDB Endowment},
% } 
% 
% 
% @inproceedings{YL87,
%  author = {Yang, H. Z. and Larson, Per-Ake},
%  title = {Query Transformation for PSJ-Queries},
%  booktitle = {Proceedings of the 13th International Conference on Very Large Data Bases},
%  series = {VLDB '87},
%  year = {1987},
%  isbn = {0-934613-46-X},
%  pages = {245--254},
%  numpages = {10},
%  url = {http://dl.acm.org/citation.cfm?id=645914.756646},
%  acmid = {756646},
%  publisher = {Morgan Kaufmann Publishers Inc.},
%  address = {San Francisco, CA, USA},
% } 
% 
% 
\noindent

%  numpages = {6},
%  url = {http://doi.acm.org/10.1145/588111.588138},
%  doi = {10.1145/588111.588138},
%  acmid = {588138},
%  publisher = {ACM},
%  address = {New York, NY, USA},
% }
% 
% 
% 
% @inproceedings{SV05,
%  author = {Segoufin, Luc and Vianu, Victor},
%  title = {Views and Queries: Determinacy and Rewriting},
%  booktitle = {Proceedings of the Twenty-fourth ACM SIGMOD-SIGACT-SIGART Symposium on Principles of Database Systems},
%  series = {PODS '05},
%  year = {2005},
%  isbn = {1-59593-062-0},
%  location = {Baltimore, Maryland},
%  pages = {49--60},
%  numpages = {12},
%  url = {http://doi.acm.org/10.1145/1065167.1065174},
%  doi = {10.1145/1065167.1065174},
%  acmid = {1065174},
%  publisher = {ACM},
%  address = {New York, NY, USA},
% } 
% 
% 
% 

%   KEYWORDS = {}}

% Using (IA) Grace can see 
% the pair\footnote{We are at level 0 now, so (IA) has four free variables, not just two. But let us try not to think about it now.} 
% $[{\bf a},{\bf a}]$ . So, as we want the two girls to see the same, Ruby must also see this pair via (IA). 
% But we cannot give her $H(\spr{}{},{\bf a},{\bf b})$.
% Instead (according to the magic of Spider Algebra), the TGD generated by (IA) will give her a new vertex $b_1$ and edges $H(\spg{}{4},{\bf a},b_1)$ and 
% $H(\spg{}{5},{\bf a},b_1)$. But now not only Ruby can see  $[{\bf a},{\bf a}]$ in (IA). She also can see the same pair in (IB). 

}

%Let now $d$ be any given natural number. To show that   
 %%%%%%%%%%%%%%%%%%%%%%%%%%%%%%%%%%%%%%%%%%%%%%%%%%%%%%%%%%%%%%%%%%%%%%%%%%%%%%%%%%%%%%%%%%%%%%%%%%%%%%%%%%%%%%%%%%%%%%%%%%%%%%%%%%%%%%%%%% 
\newpage
  
\section{Appendix A. Proof of Lemma \ref{o-kompilacji}}  

\subsection{Proof of Claim (1).}

This subsection will not be readable without reading (and understanding)  Sections IV, V and VI.A of [GM15] first.

We will introduce two operations on structures: $compile$ (turning swarms into structures) and $decompile$ (turning structures into swarms). 
They will satisfy the following properties:

 \lemmaaa{~\label{com_odpowiedniosc}
  \begin{itemize}
   \item[(i)] Let \Dd~be a swarm. If \Dd$\;\models{\cal T}$ then $compile($\Dd$)\models Compile({\cal T})$. Moreover \Dd~contains 
    a red spider (i.e. contains an atom of the relation $H(\spr{}{},\_,\_)$)
    if and only if $compile($\Dd$)$ contains a red spider (i.e. a copy of the full red spider $\spr{}{}$) and 
    \Dd~contains 
    a green spider (i.e. contains an atom of the relation $H(\spg{}{},\_,\_)$)
    if and only if $compile($\Dd$)$ contains a green spider (i.e. a copy of the full red spider $\spg{}{}$).

   \item[(ii)] Let  \Dd~be a relational structure. If \Dd$\;\models Compile({\cal T})$ then $decompile($\Dd$)\models {\cal T}$. 
   Moreover, \Dd~contains a red spider (i.e. a copy of the full red spider $\spr{}{}$) if and only if $decompile($\Dd$)$ contains a red spider
   (i.e. contains an atom of the relation $H(\spr{}{},\_,\_)$) and 
   \Dd~contains a green spider (i.e. a copy of the full green spider $\spg{}{}$) if and only if $decompile($\Dd$)$ contains a green spider
   (i.e. contains an atom of the relation $H(\spg{}{},\_,\_)$).
  \end{itemize}
}
With this Lemma proof of Lemma \ref{o-kompilacji}(1) is trivial: Existence of an appropriate model for ${\cal T}$ is equivalent to existence of such model for $Compile({\cal T})$.

\definitionnn{ The swarm
$decompile(\mathbb D)$ is defined as the
set of all  triples  $H({\cal S},b,c)$  such that
$\mathbb D\models ${\small H}$(a,b,c)$ and $a$ is the head of a real spider in $\mathbb D$ which  is isomorphic to $\cal S$.
}

This definition is natural. While ``$decompiling$'' a structure we just abstract from the physical realization 
of spider's legs. Proof of Lemma \ref {com_odpowiedniosc}(ii) is straightforward.\medskip

The second construction is a bit more involved and will require some analysis. The reason for this is that we have to figure out how the legs of the
spiders should be  connected in $compile($\Dd$)$: this information is not present in \Dd.

\definitionnn{\label{def:real}
The relational structure $compile($\Dd$)$ is defined as follows. Let \Dd$^0$ be a structure obtained by replacing each edge $H(\mathcal{S},a,b)$ of \Dd~by a real spider which is isomorphic to {$\mathcal{S}$}, with $b$ being his antenna and $a$ being his tail. We say that two knees $b_1,b_2 \in\;$\Dd$^0$ (possibly belonging to two different 
real spiders) are $\sim$-equivalent if and only if the calves connected to them have the same predicate symbol and  the same color. 
We define $compile($\Dd)$=$\Dd$^0/_\sim$.
}
  
In other words Definition \ref{def:real} says that in order to 
built  $compile($\Dd)
we create $4s$ additional vertices which are going to serve as calves ($2s$ green and $2s$ red -- one for each equivalence class of $\sim$) 
and connect each head of a spider from \Dd~ to appropriate calves.

Clearly, each edge of swarm \Dd~is represented by a real spider in $compile($\Dd$)$. 
We however need to make sure that no new real spiders emerged, apart from the ones being representations of 
edges from \Dd.
It is important, because such new spiders could serve as arguments for rules in $Complie({\cal T})$, and make the condition that 
if \Dd$\;\models{\cal T}$ then $compile($\Dd$)\models Compile({\cal T})$ unsatisfied.

\lemmaaa{\label{lem_nic_nowego}
 decompile(compile(\Dd))=\Dd
}
{\em Proof:} Let \Dd$^0$ be like in Definition \ref{def:real}.
Each head of a real spider in \Dd$^0$  is connected to exactly one thigh atom $T_j$ and exactly one thigh atom $T^j$ for each $j\leq s$. Similarly each thigh atom is connected to exactly one calf atom. Thus in \Dd$^0$ the only real spiders are those which are associated with some edge of \Dd. Therefore $decompile($\Dd$^0)=$\Dd.

Notice that each knee in $compile($\Dd$)$ is connected to exactly one calf: this was true in \Dd$^0$ and  for each  $\sim$-class $E$ the calves connected to $E$ are identical\footnote{It is important here, that all those calves share a common end, which is a constant in $\Sigma$.}. Since no new real spiders emerge in \Dd~ compared to \Dd$^0$, then  indeed $decompile(compile($\Dd$))=$\Dd.  \eop

~

\noindent
{\em Proof of Lemma \ref{com_odpowiedniosc} (i):}
Let \Dd~be a swarm  such that  \Dd$\;\models{\cal T}$. We are going to show that  $compile($\Dd$)\models Compile({\cal T})$.
  
  Suppose $s=\fff^i_j\oaa\fff^k_l\in Compile({\cal T})$   and let {\small H}$(h_1,a_1,b)$, {\small H}$(h_2,a_2,b)$ be head atoms of two real spiders $s_1,s_2$ in 
  $compile($\Dd$)$, isomorphic to the ideal spiders $\spg{i}{}$,$\spg{k}{}$ respectively (this is  of course one of many possible cases, but they are similar). We have to show that there exist two real s-piders $s_3,s_4$ in $compile($\Dd$)$ with head atoms of the form  {\small H}$(h_3,a_1,b')$, {\small H}$(h_4,a_2,b')$ isomorphic to the spiders $\spr{}{j}$,$\spr{}{l}$ respectively. 
  Moreover  spider $s_1$ must share calves upper $i$ and lower $j$ with  spider $s_3$ and  spider $s_2$ must share calves upper $k$ and lower $l$ with  spider $s_4$. 

Consider edges $H(\spg{i}{},a_1,b)$ and $H(\spg{k}{},a_2,b)$ in \Dd. They must exist due to Lemma \ref{lem_nic_nowego}. 
Since \Dd$\;\models \fff^i_j\oaA\fff^k_l$ we know that  in \Dd~must exist edges $H(\spr{}{j},a_1,b')$, $H(\spr{}{l},a_2,b')$. Let $s_3$, $s_4$ be real spiders 
in \Dd~ associated with those edges. 
Let $s_1^0$, $s_2^0$ $s_3^0$, $s_4^0$ be real spiders in \Dd$^0$ spiders  $s_1$, $s_2$ $s_3$, $s_4$ come from. 
Since $^i$-th calves of $s_1^0$ and $s_3^0$ are red, the $^i$-th knees of $s_1$ and $s_3$ are $\sim$-equivalent and hence equal in \Dd.
The same holds for  $_j$-th,$^k$-th, and $_l$-th calves.\eop

  \subsection{Proof of Claim (2)}
  
Proof of Lemma \ref{o-kompilacji}(2) has similar high level architecture as proof  of Lemma \ref{o-kompilacji}(1), but  is more complicated.

Let us first remark that it is hard to reason about {\bf arbitrary} finite models of a set of rules ${\cal T}$.
Much harder than to reason about $chase({\cal T},{\cal D})$ for some ${\cal D}$. This is because $chase({\cal T},{\cal D})$ is being built stage by stage,
and inductive argument can be very often applied. Minimal models, which we now define, retain some nice properties of $chase$.

Recall that each rule $T$ from ${\mathbb L}_1$ as well as each rule from ${\mathbb L}_2$ postulates, for  two edges of the current swarm (or current green graph) satisfying the 
left-hand side of the rule,
existence of {\em witnesses} -- two edges satisfying the right-hand side of the rule.

\definitionnn{
 Let ${\cal T}\subseteq {\mathbb L}_1$ (resp. ${\mathbb L}_2$) and let $\mathbb M$ be a model of ${\cal T}$ containing  
 ${\text \small H}(\spg{}{},{\bf a},{\bf b})$. The set of important edges of $\mathbb M$ is defined inductively. Edge $e$ is important if:
 
 \begin{itemize}
  \item $e= {\text \small H}(\spg{}{},{\bf a,b})$ 
  
  \item There exist a rule $T\in {\cal T}$  and  important edges $e_0$ and $e_1$, which satisfy the left-hand side of $T$, such that $e$ and some other $e'$ in $\mathbb M$
  are the postulated pair of witnesses. 
 \end{itemize}

 We say that $\mathbb M$ is a minimal model of ${\cal T}$ if $\mathbb M$ contains the edge $H(\spg{}{},{\bf a,b})$
 and  every edge in $\mathbb M$ is important.
}

It is clear that if there exists a model, then there exists a minimal one. One can just take a substructure containing only important edges as a new model.
Notice that since importance of edges is defined in an inductive way, we can now use induction to prove our lemmas.

Now -- for a fixed set $\cal T$ of green graph rewriting rules -- 
we define two more operations on structures: $precompile$ (turning green graphs being {\bf  minimal models} of ${\cal T}$ into swarms) and $deprecomplie$
(turning swarms being {\bf  minimal models} of ${\cal T}$ into green graphs). They will satisfy the following properties:

 \lemmaaa{~\label{precom_odpowiedniosc}
  \begin{itemize}
   \item[(i)] Let  \Dd~be a swarm without (an edge labelled with)  full red spider, a minimal model of $Precompile(\Tt)$. Then $deprecompile(\Ddd)\models \Tt$. Moreover $deprecompile(\Ddd)$ contains no 1-2 pattern.
   \item[(ii)] Let \Dd~be a green graph without 1-2 pattern, a minimal model of $\Tt$. Then $precompile(\Ddd)\models Precompile(\Tt)$. Moreover  
   $precompile(\Ddd)$  contains no (edge labelled with) full red spider.
  \end{itemize}
}

\noindent
{\bf Operation {\em deprecompile}.}
Let us remind the reader that spider $\spg{I}{J}$ (or $\spr{I}{J}$) is called {\em lower} if $J$ is non-empty. In a similar manner:

\definitionnn{Rule $\fff^{I_1}_{J_1}\obB \fff^{I_2}_{J_2}$ (or $\fff^{I_1}_{J_1}\oaA \fff^{I_2}_{J_2}$) from ${\mathbb L}_1$ is {\em lower} if both 
$J_1$ and $J_2$ are non-empty.
}

The proof of the following lemma is by (easy) induction, in the spirit of the proof of [GM15, Lemma 16]:

\lemmaaa{\label{czerwonedolne}
Let ${\cal T}\subseteq {\mathbb L}_1$ be a set of  lower rules and let swarm $\mathbb M$ be a minimal model of ${\cal T}$.
%, without (an edge labelled with) 
%full red spider. 
Let  ${\cal S}$ be the spider 
being the label of some edge in $M$. Then  ${\cal S}$ is red if and only if it is lower. %In particular  $\spr{}{}$ is not a label of any edge in $M$.
}

\definitionnn{For a swarm $\mathbb D$, the green graph $deprecompile(\Ddd)$ is the set of full or upper 1-lame green edges from \Dd.}

In other words, $deprecompile(\Ddd)$ is what remains of swarm $\Ddd$ after removing everything that is not a valid edge of a green graph.

\noindent
{\em Proof of Lemma \ref{precom_odpowiedniosc}(i):} The green graph $deprecompile(\Ddd)$ contains no 1-2 pattern. 
Otherwise application of the sequence of rules  $\fff^{1} _1\oaA \fff^{2}_2$,~ $\fff^{3}_1 \oaA \fff^{4}_2$,~$\fff^{3} \oaA \fff^{4}_3$ 
to this 1-2 pattern would have created a full red spider in $\mathbb D$. 

Each rule in $\Tt$ can be seen as a composition of two rules in $Precompile(\Tt)$: if a rule $T$ from $\Tt$ postulates existence of some witnesses,
then there exists a pair of rules $T_1, T_2$ in $Precompile(\Tt)$ which  implies existence of the same witnesses. Since \Dd~is a model of $Precompile(\Tt)$ it follows that $deprecompile(\Ddd)$ is a model of each composition mentioned before. Therefore  $deprecompile(\Ddd)\models \Tt$.\eop\medskip

\noindent
{\bf Operation {\em precompile}.} To transform a green graph into a swarm we have to at least add some red edges, since  rules in
$\mathbb{L}_1$ require witnesses in color opposite to the color of arguments. As it turns out, it is enough to add red edges produced by a single stage of $chase$:

% 
% \definitionnn{Let $s\in \mathbb{L}_1$ and let $g_1,g_2$ be a pair of edges such that it satisfy left-hand side of $s$. By $s(g_1,g_2)$ we denote the postulated pair of edges. If $s$ is a rule of type $\obB$ (resp. $\oaA$), then edges in $s(g_1,g_2)$ share tails (resp. antennas) with $g_1,g_2$, whereas common antenna  (resp. tail) of $s(g_1,g_2)$ is a new vertex. }
% 
% \definitionnn{By $W'$ we denote the following superstructure of \Dd~containing new witnesses for rules in $Precompile({\cal T})$:  $$W'= \Ddd \cup \{s(g_1,g_2)| s\in Precompile({\cal T}),~ g_1,g_2\in \Ddd \}$$  and assuming that ${\text \small H}(\spg{}{},a,b) \in \Ddd$ let
% $$W= W' \cup \{{\text \small H}(\spg{3}{i},a,b_{ij}), {\text \small H}(\spg{3}{j},a,b_{ij}) | \fff_i \text{$\oaA$} \fff_j \in Precompile({\cal T})\}$$ where $b_{ij}$ are new vertexes. Let  $Precompile(\Ddd) \subseteq W$ be a minimal model for $Precompile(\Tt)$.}
%   
\definitionnn{
For a green graph $\mathbb D$, a minimal model of $\cal T$, let $precompile({\mathbb D})$ be the swarm
$chase_1(Precompile(\cal T),{\mathbb D})$.
}

In other words, $precompile({\mathbb D})$ is $\mathbb D$ plus all the red edges demanded, as witnesses,  by the rules in $Precompile(\cal T)$ with arguments
from $\mathbb D$. Notice that no green edges are added.

\lemmaaa{
Suppose $\mathbb D$ is a green graph like in Lemma \ref{precom_odpowiedniosc}(ii). Then no edge in $\mathbb D$ is labelled with $\spg{3}{}$. 
}

\noindent
{\em Proof:} There is no rule in $\cal T$ involving $\spg{3}{}$, so there is no way for such edge to be important.\eop\medskip

\noindent
{\em Proof of Lemma \ref{precom_odpowiedniosc}(ii):} 

First let us show that there is no edge labelled with $\spr{}{}$ in $precompile({\mathbb D})$. The only rule which is not lower in $Precompile(\cal T)$
is  $\fff^{3} \oaA \fff^{4}_3$, and it is the only one which could produce  $\spr{}{}$. But it would need an edge labelled with $\spg{3}{}$ as one of the arguments, 
and, as we have already noticed, there is no such edge in $\mathbb D$.

It remains to show that $precompile({\mathbb D})$ is indeed a model of $Precompile(\cal T)$. 

Clearly, there are needed (red) witnesses for all the rules in  $Precompile(\cal T)$ with pairs of green edges as arguments. 
We need to prove that this is also the case for pairs of red edges as arguments. 

Let $e_1$ and $e_2$ be two edges 
in $\mathbb D$
sharing  antennas (or tails, the argument is analogous), labelled with green spiders ${\cal S}_1$ and ${\cal S}_2$
and let  $e'_1$ and $e'_2$, labelled with ${\cal S}'_1$ and ${\cal S}'_2$
be red edges of $precompile({\mathbb D})$ added by some rule $\fff^{I_1}_{J_1}\oaA \fff^{I_2}_{J_2}$ having
$e_1$ and $e_2$,  as arguments. Edges $e'_1$ and $e'_2$ also share antennas, and $e_1$ shares its tail with $e'_1$ while 
$e_2$ shares its tail with $e'_2$.

We want to show that whenever $e'_1$ or $e'_2$ is an argument of some rule, then 
the demanded (green) witness was already in $\mathbb D$ (and thus is in  $precompile({\mathbb D})$)
 
There are several cases, none of them difficult. As an example we will consider one of them, analysis is similar for all other cases:

Suppose that none of ${\cal S}_1$ and ${\cal S}_2$ is  $\spg{}{}$. This means (according to the Rule of Spiders Algebra $\clubsuit$)
that ${\cal S}_1= \spg{I_1}{}$ and ${\cal S}_2=\spg{I_2}{}$, with $I_1,I_2\neq\emptyset$. Notice that 
$\fff^{I_1}_{J_1}\oaA \fff^{I_2}_{J_2}$ is neither $\fff^{1}_{2}\oaA \fff^{2}_{2}$ (because there is no 1-2 pattern in $\mathbb D$) 
nor $\fff^{3}_{1}\oaA \fff^{4}_{2}$ (because there is no edge labelled with $\spg{3}{}$). 
Using  $\clubsuit$ again we get that ${\cal S}'_1=\spr{}{J_1}$ and ${\cal S}'_2=\spr{}{J_2}$.
There are only two rules in $Precompile(\cal T)$ which match with any of $\spr{}{J_1}$ or $\spr{}{J_2}$:
one of them is $\fff^{I_1}_{J_1}\oaA \fff^{I_2}_{J_2}$ and another is some $\fff^{I'_1}_{J_1}\oaA \fff^{I'_2}_{J_2}$, with
$I_1\oAA I_2 \leftrightarrowtriangle I'_1\oAA I'_2 $ being one of the green graph rewriting rules of $\cal T$. Since 
$\mathbb D$ is a model of $\cal T$ we know that there are edges $e''_1$ and $e''_2$ in $\mathbb D$, sharing antennas and 
such that and $e''_1$ shares its tail with $e'_1$ while 
$e''_2$ shares its tail with $e'_2$, with $\spg{I'_1}{}$ being the label of $e''_1$ and $\spg{I'_2}{}$ being the label of $e''_2$.
Now $e_1$ with $e_2$ and $e''_1$ with $e''_2$ are pairs of witnesses for both possible applications of rules involving $e'_1$ or $e'_2$. 

Notice that we (silently) used the (easy) observation here, that the joint antenna of $e'_1$ and $e'_2$ does not belong to any other edge, 
and so the only edge that can be used as an argument in a rule of the type ``$\oaA$'' together with $e'_1$ is $e'_2$ (and the other way round).
\eop

\section{Appendix B. Proof of Lemma \ref{tugranice}}

\noindent
Clearly, ${\mathbb M}\models \sep_\infty$, so for the proof of Lemma \ref{tugranice} (ii) we only need to show that ${\mathbb M}\models \sep_\boxplus$.

Call the edges of $\mathbb M$ which are labelled with $\spg{}{}$, $\spg{\alpha}{}$, $\spg{\beta_0}{}$, $\spg{\beta_1}{}$, $\spg{\eta_0}{}$, or $\spg{\eta_1}{}$
{\em skeleton}, and  the  edges of $\mathbb M$ which are labelled with any of the $\spg{\langle \_,\_,\_,\_\rangle}{}$ {\em foam}. Among the foam 
edges we will distinguish between ${\bar b}$-foam  and  $b$-foam (depending whether the label is of the form 
$\spg{\langle \_,\_,\_,\bar b\rangle}{}$ or of the form $\spg{\langle \_,\_,\_,b\rangle}{}$). We will also refer to 
north-edges, which are foam edges of the form  $\spg{\langle n,\_,\_,\_\rangle}{}$.

The next lemma follows directly from the construction of $\mathbb M$: 

\lemmaaa{\label{foamlemma}
\begin{enumerate}
\item
If $e_1$ and $e_2$ are two edges of $\mathbb M$ which share at least one vertex, and such that $e_1$ is a skeleton edge, and $e_2$ is a foam edge,
then $e_2$ is $b$-foam.
\item
If $e_1$ and $e_2$ are two edges of $\mathbb M$ which share at least one vertex, and such that $e_1$ is a skeleton edge, and $e_2$ is a foam edge from some 
${\mathbb M}_t$ then $e_1$ is also an edge of ${\mathbb M}_t$.
\item
If $e$ and $e'$ are two foam edges of $\mathbb M$, $e$ coming from ${\mathbb M}_t$ and $e'$ coming from ${\mathbb M}_{t'}$, for some $t\neq t'$, which 
share at least one vertex, then they both share this vertex with some skeleton edge (and hence are -- due to (1) -- are both $b$-foam).
\item 
If $e$ is a north-edge of some ${\mathbb M}_t$ and $e$ shares its end with some edge $e'$ in $\mathbb M$ then $e'$ is also an edge of ${\mathbb M}_t$.
\end{enumerate}

}

Now we are ready to prove claim (1) of Lemma \ref{tugranice}. Clearly, there is no 1-2 pattern in any of the ${\mathbb M}_t$. So the only 
way to have it in $\mathbb M$ would be if (*) one of the edges of the 1-2 pattern came from ${\mathbb M}_t$ and another from ${\mathbb M}_{t'}$, for  some $t\neq t'$.
But the two edges of a 1-2 pattern (**) are $\bar b$-foam edges which share a vertex. By Lemma \ref{foamlemma}(3) conditions (*) and (**) are contradictory.   

Concerning Lemma \ref{tugranice}(2), for each $t\in \mathbb N$ there is ${\mathbb M}_t\models \sep_\boxplus$.
So again, like in the proof of Lemma \ref{tugranice}(1) 
the 
only way that ${\mathbb M}\not\models \sep$ could possibly happen would be if we were able to find two edges $e$ and $e'$ which share a vertex, do not come
from the same  ${\mathbb M}_t$, and such that some rule from $\sep_\boxplus$ applies to them. By Lemma \ref{foamlemma}(2) none of them is a skeleton edge. By Lemma \ref{foamlemma}(3) they are both $b$-foam edges. The only rule of $\sep$ which can rewrite two  $b$-foam edges is the grid triggering rule, but 
this would require $e$ and $e'$ to share their end-vertices and  one of them to be a north-edge. Which is impossible due to  Lemma \ref{foamlemma}(4).\eop

\section{Appendix C. Proof of Lemma \ref{bezsenny}}  
  
By an ${\bf ab}$-path in a green graph $\cal D$
we will mean a directed path in $PG({\cal D})$ leading from ${\bf a}$ to ${\bf a}$ or to ${\bf b}$. By a $\mathfrak Q$-edge we will mean  an edge labelled with an element of $\mathfrak Q$.
%(in a graph whose edges are labelled with elements of $\mathfrak Q$ and $\mathfrak A$).

A pair 
$H_{{\mathfrak c'}}(c,d'), H_{{\mathfrak d'}}(c',d')$ [or $ H_{{\mathfrak c'}}(d',c), H_{\mathfrak d'}(d',c')$] of edges 
of vertices of a green graph which satisfies condition $\spadesuit$ for some rule will be called {\em a right-match}. 
If also  condition $\heartsuit$ is  satisfied for  $c, c'$ then we say that they are {\em an interesting right-match}. Analogously we define 
{\em left-match}, and {\em interesting left-match}. 

Notice that a green graph is a model of $\sep_\vartriangle$ if and only if there are no interesting matches there.
Also notice that, at each of its elementary steps\footnote{By ``elementary step'' we mean a single execution of the inner loop.}, the {\bf procedure} which constructs $\mathbb M$ finds an interesting right-match
 and
adds a vertex and two edges (or just one edge) which match with the left-hand side of the same rule. 
Well, it is indeed clear that this is what an elementary step does if ${\mathfrak d}\neq\emptyset$.
To see that it is also the case when
 ${\mathfrak d}=\emptyset$ we need:
 \lemmaaa{
 \textbullet ~ If ${\mathbb M}^m\models H_{\mathfrak a}({\bf a},c)$ then ${\mathfrak a}\in \{\alpha,\eta_{11},\eta_1 \}$\\
 \textbullet ~ If ${\mathbb M}^m\models H_{\mathfrak a}(c,{\bf b})$ then ${\mathfrak a}\in \{\eta_0, \omega_0 \}$\\
 
 \textbullet ~ If ${\mathbb M}^m\models H_{\mathfrak a}(d,c)$ and ${\mathfrak a}\in \{\alpha,\eta_{11},\eta_1 \}$ then $d={\bf a}$.\\
 \textbullet ~ If ${\mathbb M}^m\models H_{\mathfrak a}(c,d)$ and ${\mathfrak a}\in \{\eta_0, \omega_0 \}$ then $d={\bf b}$.
 }
 
 Proof of this Lemma is by straightforward induction/case inspection. 
 
 Now one can easily notice that when an elementary step
 is executed, in the situation when ${\mathfrak d}=\emptyset$, then $H_\emptyset({\bf a},{\bf b})$, together with the newly added edge, form the demanded
 left-match.

Since neither $\beta_0$ nor $\beta_1$ ever occur
in the left-hand side, the second claim of Lemma \ref{bezsenny} is obvious.
The first claim may seem to be less straightforward. If we only use the rules of $\sep_\vartriangle$ in the right-to-left direction, how can we be sure that 
the resulting green graph has no interesting left-matches? And also, 
if we terminate the {\bf procedure} after a fixed number  ($k_\vartriangle$) of steps
how can we be sure there are no interesting right-matches left?

It is straightforward to prove by induction that for each $ 0\leq m\leq k_\vartriangle+1$:

\lemmaaa{[loop invariants]\label{struktura-thuego} 
\begin{enumerate}
\item
If ${\mathfrak v}\in words({\mathbb M}^m)$ then ${\mathfrak v} \stackrel{\ast\;\;}{\rightsquigarrow_\vartriangle} {\mathfrak u}_\vartriangle$. 
\item 
Every path in ${\mathbb M}^m$ is a subpath of 
some $\bf ab$-path  in $PG({\mathbb M}^m)$. In particular, every  edge of ${\mathbb M}^m$ belongs to   some $\bf ab$-path  in $PG({\mathbb M}^m)$.
\item 
 On each $\bf ab$-path  in $PG({\mathbb M}^m)$ there is exactly one $\mathfrak Q$-edge.
\item
Every $\mathfrak Q$-edge in $PG({\mathbb M}^m)$ is on exactly one  $\bf ab$-path in $PG({\mathbb M}^m)$.
\end{enumerate}
}

For the induction step it is crucial to realize that if a rule of ${\mathfrak T}_\vartriangle$ is of the form 
$ {\mathfrak c}\oBB {\mathfrak d} \leftrightarrowtriangle  {\mathfrak c'}\oBB {\mathfrak d'} $ then 
${\mathfrak c}$ and ${\mathfrak c}'$ are odd and ${\mathfrak d}$ and ${\mathfrak d}'$ are even and if 
some rule of ${\mathfrak T}_\vartriangle$ is of the form 
$ {\mathfrak c}\oAA {\mathfrak d} \leftrightarrowtriangle  {\mathfrak c'}\oAA {\mathfrak d'} $ then 
${\mathfrak c}$ and ${\mathfrak c}'$ are even and ${\mathfrak d}$ and ${\mathfrak d}'$ are odd.
In consequence, in one elementary step of the {\bf procedure} a path of length 2 (or 1), from some $c$ to $c'$ is added in 
$PG({\mathbb M}^{m+1})$ only if already in $PG({\mathbb M}^{m})$ there already was a path of length 2 from  $c$ to $c'$

Since -- as it follows from Lemma \ref{struktura-thuego}(4) -- each $\mathfrak Q$-edge $e$ of $PG({\mathbb M}^m)$ 
 is on exactly one $\bf ab$-path in $PG({\mathbb M}^m)$, and we know that this path 
can be read as a word, we can call this word ${\mathfrak w}(e)$.

\lemmaaa{\label{zarazzaraz}
There are no interesting left-matches in ${\mathbb M}^m$

%   For each rule 
%   $ {\mathfrak c}\oAA {\mathfrak d} \leftrightarrowtriangle  {\mathfrak c'}\oAA {\mathfrak d'} $ 
% [or $ {\mathfrak c}\oBB {\mathfrak d}\leftrightarrowtriangle  {\mathfrak c'}\oBB {\mathfrak d'} $] in $\sep_\vartriangle$
%   and for each triple  $c,c',d$ of verticies of ${\mathfrak M}^m$ such that 
%   ${\mathbb M}^m\models H_{{\mathfrak c}}(c,d), H_{{\mathfrak d}}(c',d)$
%   [or  ${\mathbb M}^m\models H_{{\mathfrak c}}(d,c), H_{\mathfrak d}(d,c')$]
%   there exists $d'$ such that 
%   ${\mathbb M}^m\models H_{{\mathfrak c'}}(c,d'), H_{{\mathfrak d'}}(c',d')$
%   [or  ${\mathbb M}^m\models H_{{\mathfrak c'}}(d',c), H_{\mathfrak d'}(d',c')$].
  }
 
 \noindent
{\em Proof:} This is of course true in ${\mathbb M}^0$, since ${\mathfrak u}_\vartriangle$ is the configuration in which $\Delta$ terminates. 
For the induction step imagine a  vertex and two  edges $e$ and $e'$ (or just one edge, the argument is analogous) 
were added to the current structure, as the result of one of the elementary steps. 
It follows from the form of the rules of $\sep_\vartriangle$ that one of them -- let it be $e$ -- is a $\mathfrak Q$-edge. Rainworms are deterministic,
so there exists at most one rule of $\Delta$ which can be applied to  ${\mathfrak w}(e)$, and thus there is at most one rule in  $\sep_\vartriangle$ whose
left-hand side matches with some edges on the $\bf ab$-path the edge $e$ belongs to. But we already know one such rule -- it is the one that created $e$ and $e'$,
which means that the edges that match with its right-hand side must have existed in the structure before $e$ and $e'$ were created. \eop

\lemmaaa{\label{jest-skonczony}For each $m\leq k_\vartriangle$ the structure ${\mathbb M}^m$ is finite.
}
       
\noindent 
{\em Proof:} ${\mathbb M}^0$ is finite. If ${\mathbb M}^m$ is finite then there is a finite number of $\mathfrak Q$-edges there, and for each of them there 
are at most $c_\vartriangle$ right-matches, so ${\mathbb M}^{m+1}$ is also finite.\eop\medskip

It follows from Lemma \ref{struktura-thuego}(1) that  if $e$ is a $\mathfrak Q$-edge of $\mathfrak M$ then
       ${\mathfrak w}(e) \stackrel{\ast\;\;}{\rightsquigarrow_\vartriangle} {\mathfrak u}_\vartriangle$ and, in consequence 
       (from Lemma \ref{jakis-porzadek-2}(1) and (3)), we have 
       ${\mathfrak w}(e) \stackrel{k_e\;\;}{ \rightsquigarrow_\vartriangle} {\mathfrak u}_\vartriangle$ for some natural number $k_e\leq k_\vartriangle$.
       
       On the other hand, there of course exists, for each edge $e$ in $PG({\mathbb M})$, the minimal number $m$ such that
       $e$ is already in $PG({\mathbb M}^m)$. This minimal $m$ will be denoted as $m_e$. 
       Last lemma we need for the proof of Lemma \ref{bezsenny} is:
       
       \lemmaaa{\label{sliczny}
       Let $e$ be a $\mathfrak Q$-edge of $PG({\mathbb M}^m)$. Then $m_e=k_e$.}

It follows from the form of the rules in $\sep_\vartriangle$ that each edge of $\mathbb M$ must have been created together with some 
$\mathfrak Q$-edge. So once Lemma \ref{sliczny} is proven we  know  that 
${\mathbb M}={\mathbb M}^{k_\vartriangle+1}=M^{k_\vartriangle}$, 
which means that the last execution of the main loop in the procedure defining ${\mathbb M}$ was in vain --
 there were no interesting right-matches in $M^{k_\vartriangle}$ any more.
 This -- together with Lemma \ref{zarazzaraz} and Lemma \ref{jest-skonczony} will imply that ${\mathbb M}\models \sep_\vartriangle$.\\
 
 \noindent
 {\em Proof of Lemma \ref{sliczny}:}
 The claim is of course true for $m=0$. 
 
 For the induction step first notice that if edges $e$ and $e'$ are a right-match in ${\mathbb M}^m$ and $m'>m$ then 
 $e$ and $e'$ do not form an interesting  right-match in ${\mathbb M}^{m'}$ -- this is since condition $\heartsuit$ was no longer satisfied for them
 in ${\mathbb M}^{m+1}$ at latest. Notice also that if edges 
 $e$ and $e'$ are a right-match in ${\mathbb M}$ and $e$ is a $\mathfrak Q$-edge then $e'$ exists already in ${\mathbb M}^{m_e}$ (this is because
 $e$ only belongs to one $\bf ab$-path in ${\mathbb M}$ and this path exists already in ${\mathbb M}^{m_e}$  since we
 know $e$ belongs to some  $\bf ab$-path already in ${\mathbb M}^{m_e}$). In consequence, 
 a $\mathfrak Q$-edge $e$ can be a part of an interesting right-match only in ${\mathbb M}^{m_e}$. 
 
 Assume now the claim is true for some $m$ and let $e_1$ be a  $\mathfrak Q$-edge such that $m_{e_1}=m+1$. This means there was an 
 interesting right match 
 in ${\mathbb M}^{m}$, consisting of some $e$ and $e'$, with $e$ being a  $\mathfrak Q$-edge, such that $e_1$ was created (possibly together with some 
 $e_1'$) in some elementary step using this right match. Of course (this is how the rules of ${\mathfrak T}_\vartriangle$ work)
 we have ${\mathfrak w}(e_1) {\rightsquigarrow_\vartriangle} {\mathfrak w}(e)$, so $k_{e_1}=k_e +1$. From the reasoning in the previous paragraph 
 we get that $m=m_e$ and by hypothesis,
 we have that $m_e=k_e$. So  $k_{e_1}=m+1=m_{e_1}$.\eop

\end{document}